\documentclass[journal=acs,manuscript=article]{achemso}
\usepackage[utf8]{inputenc}
\DeclareUnicodeCharacter{2212}{-}

\usepackage{amssymb}
\usepackage{chemformula}
\usepackage{braket}

\newcommand{\parencite}[1]{\citenum{#1}} \newcommand{\textcite}[1]{\citet{#1}} 

\usepackage[colorlinks=true, allcolors=blue]{hyperref}
\usepackage[T1]{fontenc}
\usepackage{textgreek}

\title{Revealing antenna-molecule coupling using virtual gain from synthetic complex frequency analysis}

\author{Toby Severs Millard}
\affiliation{Blackett Laboratory, Department of Physics, Imperial College London, London SW7 2AZ, UK}
\alsoaffiliation{Quantum Electrical and Material Metrology Group, Quantum Technologies Department, National Physical Laboratory, Teddington TW11 0LW, UK}
\email{t.severs-millard21@imperial.ac.uk}

\author{Xiaofei Xiao}
\affiliation{Blackett Laboratory, Department of Physics, Imperial College London, London SW7 2AZ, UK}

\author{Mayela Romero-Gómez}
\affiliation{Quantum Electrical and Material Metrology Group, Quantum Technologies Department, National Physical Laboratory, Teddington TW11 0LW, UK}

\author{Tom Aktaş-Vincent}
\affiliation{Quantum Electrical and Material Metrology Group, Quantum Technologies Department, National Physical Laboratory, Teddington TW11 0LW, UK}
\alsoaffiliation{Infinitesima Ltd, Abingdon OX14 1RG, UK}

\author{Nathaniel J. Huáng}
\affiliation{Quantum Electrical and Material Metrology Group, Quantum Technologies Department, National Physical Laboratory, Teddington TW11 0LW, UK}

\author{Chris C. Phillips}
\affiliation{Blackett Laboratory, Department of Physics, Imperial College London, London SW7 2AZ, UK}

\author{Stefan A. Maier}
\affiliation{Blackett Laboratory, Department of Physics, Imperial College London, London SW7 2AZ, UK}
\alsoaffiliation{
School of Physics and Astronomy, Monash University, VIC 3800, Australia}

\author{Olga Kazakova}
\affiliation{Quantum Electrical and Material Metrology Group, Quantum Technologies Department, National Physical Laboratory, Teddington TW11 0LW, UK}

\author{Rupert F. Oulton}
\affiliation{Blackett Laboratory, Department of Physics, Imperial College London, London SW7 2AZ, UK}
\email{r.oulton@imperial.ac.uk}

\begin{document}

\begin{abstract}
Synthetic complex frequency waves (CFWs) enhance spectral and spatial resolution by compensating loss through virtual gain. This approach is particularly useful in nano-photonics where loss is intrinsically linked to strong field confinement and system size. Here, we apply CFWs to metasurfaces composed of Fano-resonant plasmonic antennas with lithographically tunable coupling between the weak and strong coupling regimes and where loss may be reduced to the Drude damping limit. We quantitatively assess the impact of the virtual gain to find that, while reducing loss, internal coupling rates remain invariant. While the exceptional point associated with strong light-matter coupling is invariant under virtual gain, the cooperativity of Fano interference is enhanced. When the metasurfaces are functionalised with a molecular monolayer, CFWs are thus able to reveal a molecule's vibrational modes, even when obscured by loss, and allow direct extraction of the plasmon-molecule coupling strength. We thus use CFW analysis as a quantitative tool for revealing and assessing light matter-coupling.

\end{abstract}

\maketitle

\section{Introduction}
The sensitivity of spectroscopic measurements of an analyte stems from the interaction between electromagnetic (EM) radiation and matter, via its electronic and vibrational states, enabling qualitative and quantitative access to a system’s composition, structure, and dynamics. This sensitivity underpins the widespread use of spectroscopy across diverse fields, from \textit{in vivo} medical diagnostics,\cite{Kim2020_Spec_vivo_medicalDiag} to determining the atmospheric composition of exoplanets,\cite{Tinetti2012_WaterExoPlanets} as well as monitoring pollutants in water.\cite{Michel2004_spec_wastewater}
The prominence of any spectroscopic signal is governed by the coupling strength between the oscillating EM field and the system of interest, as well as by their associated loss rates. Weak light-matter interactions yield low signal levels, while high losses mask spectral features behind a broadened response.

Light-matter coupling can be substantially enhanced by local EM field confinement within subwavelength volumes via the coherent collective oscillation of free electrons at metal-dielectric interfaces, known as surface plasmons.\cite{Barnes2003_plasFunda,Ciraci2012_NPoM_limOfPlasEnhc,Chikkaraddy2016_SingleMol_Plas_SC}
This gives plasmonic nanostructures their excellent sensitivity enabling a wide range of applications, including chemical vapour and gas detection,\cite{Vaidyanathan2024_PlasGasSens} single-molecule spectroscopy,\cite{Taylor2017_SingleMolePlas_rev} biomedical imaging,\cite{Mogera2022_PlasSweatAnalysis} and catalytic reaction monitoring.\cite{Zhang2015_PlasCatalMonitorREV,Stewart2008_plasSensRev,Qilin2021_PlasSensRev,xiao2021ultrabroad,schulz2024roadmap}
These sensing capabilities rely on both incoherent techniques, such as surface-enhanced fluorescence\cite{Fort2008_SEFrev} and Raman spectroscopy,\cite{SHARMA2012_SERS_overview} and coherent approaches, including refractive index sensing\cite{Xu2019_Plas&Phot_RIsensREV} and surface-enhanced infrared absorption (SEIRA).\cite{Yang2018_SEIRArev}
Despite these advantages, surface plasmons experience substantial loss.\cite{Boriskina2017_PlasmonicLossReview}
Losses may be partially reduced by incorporating gain media\cite{DeLuca2011_GainAsistedPlas} or by employing alternative plasmonic materials, including alkali metals,\cite{Qu2025_PlasAlkMetals} two-dimensional materials\cite{xiao2018theoretical,xiao2021ultrabroad}, and heavily doped semiconductors.\cite{Gururaj2013_AlternativePlasMat}
However, gain integration alters the underlying optical mechanism and can complicate sensing operation, while alternative materials introduce challenges such as high chemical reactivity\cite{Hopper2022_challenges_of_altPlas}, increased fabrication complexity\cite{Yuan2023_plas2Dsensors,Wang2020_NaPlas_fab} and difficult functionalisation processes\cite{Georgakilas2012_grapheneFuncitonalisation} relative to noble metals.
More fundamentally, the trade-off between EM confinement and loss is intrinsic to plasmonic systems, rendering light-matter coupling strength and dissipation inherently interdependent, regardless of the material platform.

Complex-frequency wave (CFW) analysis has recently been developed to retroactively reduce the effects of loss in a spectroscopic signal. The technique applies synthetic temporally attenuated waves that mimic the effect of gain and thus compensate for loss in a coherent measurement. To date, synthetic CFWs have been used to reduce loss in a metallic superlens,\cite{Guan2023_CFWSupLens} enhance the sensitivity of a graphene-based SEIRA molecular sensor\cite{Zeng2024_CFW_MoleSens} and artificially increase the propagation length of surface plasmon polaritons on hBN and \ch{MoO3}.\cite{Guan2024_CFW_SPP} While these studies demonstrate the power of synthetic CFW illumination through qualitative investigation, here we explore the quantitative capabilities of the approach. We apply synthetic CFWs to MIR multi-mode plasmonic antennas exhibiting Fano interference to measure accurately their internal coupling strength and loss. We find that the coupling strength and the transition from weak to strong coupling, known as the exceptional point, remain invariant to the virtual gain, and so may be measured from the CFW analysis. However, if the virtual gain is too high, the resulting spectral response departs from the analytical model, which reduces confidence in the extracted parameters. Typically, only half of the system loss may be compensated with confidence. We apply these findings to determine the coupling strength of a monolayer of BPT molecules to the antennas. We are thus able to extract an accurate coupling strength between the antenna and molecular vibrational modes that were initially obscured by loss.

\section{Results and discussion}
\subsection{Loss compensation in a coupled system}
\begin{figure*}[ht!]
    \centering
    \includegraphics[width=0.99\linewidth]{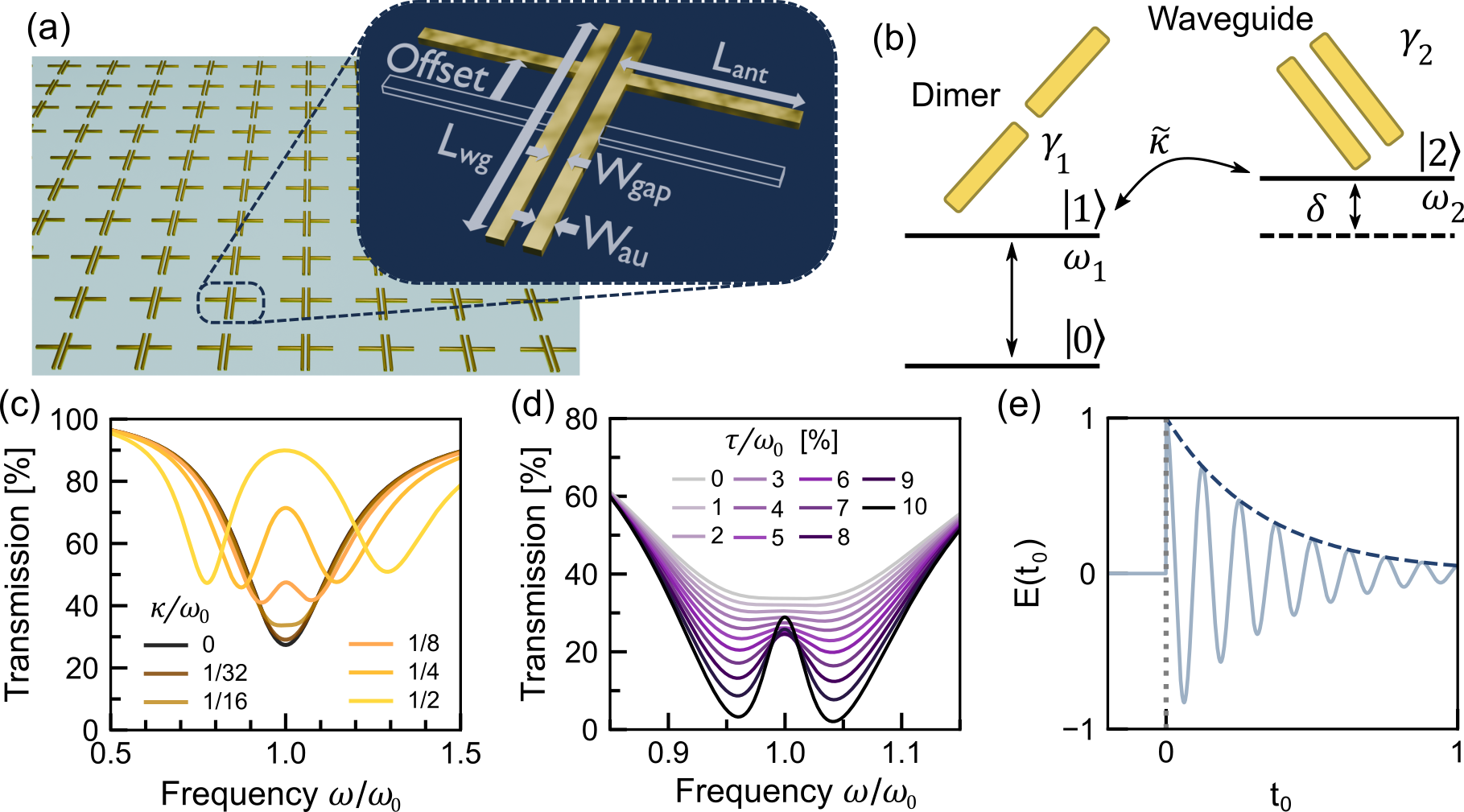}
    \caption{\textbf{The resonant dimer-waveguide nanostructure, coupled-mode theory and CFWs}. (a) Graphical illustration of the metasurface arrays with a single nanostructure inset. The labelled dimensions define the optical coupling. (b) An energy level diagram describing the coupling between constituent resonators of the nanostructure. (c) The transmission intensity of metasurfaces at various degrees of coupling, calculated using coupled-mode theory. (d) Calculated transmission of the $\kappa/\omega_0=1/16$ metasurface with loss reduced by up to $\tau/\omega_0=10\%$. (e) Amplitude of an attenuated and truncated CFW. The purple dashed line indicates the attenuation due to the imaginary component, $\tau$. }
   
    \label{fig1_CFWintro}
\end{figure*}

We study the effect of synthetic CFWs on plasmonic metasurface arrays, illustrated in Fig.~\ref{fig1_CFWintro}(a). Each antenna consists of a dark resonator of two parallel gold wires, which we call the waveguide antenna, and a bright resonant dimer antenna arranged perpendicular to the waveguide, as shown in the inset of Fig.~\ref{fig1_CFWintro}(a). 
The dimensions of the compound antenna define the constituent resonant modes and the waveguide length ($L_{wg}$), antenna length ($L_{ant}$) and waveguide gap ($W_{wg}$) are iteratively tuned using numerical simulations to achieve degeneracy at 10 \textmu m (Sec.~SI of the Supporting Information (SI)).
The resonance selected provides a spectral range that overlaps with many IR active molecules for potential functionalisation.\cite{Benda2022_MoleVibrationDB}
Fig.~\ref{fig1_CFWintro}(b) shows an energy level diagram for the system, where the resonant frequencies of the dimer and waveguide are $\omega_{1,2}$, their loss rates are $\gamma_{1,2}$ and $\delta=\omega_1-\omega_2$ represents any undesirable spectral detuning.
The dimer's bright mode is excited by far-field illumination polarised along its longitudinal axis, which subsequently couples the EM field into the waveguide gap mode through the near field, $\ket{0}\rightarrow\ket{1}\rightarrow\ket{2}$.
This coupling is represented by the complex coupling $\widetilde\kappa=\kappa e^{i\phi_k}$, where $\kappa$ is the coupling amplitude and the phase, $\phi_k$, accounts for the retardation between the two resonators.\cite{Taubert_2013_complexPhi_CMtheory}
The resonant surface plasmon polariton (SPP) modes of the waveguide are inherently nonradiative and, reciprocally, cannot be excited directly from the far field, $\ket{0} \nrightarrow \ket{2}$.
The resonant antisymmetric SPPs modes have antinodes of high EM field amplitude at each end of the waveguide. 
For the fundamental mode, a node with zero field amplitude is positioned at the central point along the waveguide.
Therefore, offsetting the dimer position along the waveguide from the central point, as shown in the inset of Fig.~\ref{fig1_CFWintro}(a), increases the coupling, as the overlap between the dimer's concentrated field and the bright region of the waveguide mode increases.\cite{Liu2009_EIT?_plasDerudeLim,Xiao2024_nanofocus}
Numerical simulations and experimental verification of the near-field coupling mechanism can be found in Sec.~SII of the SI.
We define this offset as a percentage along half the waveguide length ($L_{wg}/2$) from its centre, accounting for the waveguide's two-fold symmetry.
The subwavelength waveguide gap, $W_{gap}=100$ nm, acts to confine the resonant SPP mode to a greater degree than is achievable with a single interface, enhancing the EM field amplitude.

The system can be modelled using coupled-mode theory, where the compound plasmonic nanoresonator is treated as a system of coupled harmonic oscillators.\cite{Xiao2024_nanofocus}
Following the derivation in Sec.~SIII.I of the SI, a coupling matrix is found to describe the system.
\begin{equation}  
\label{EQ_matrix_CoupledHarm}
\begin{bmatrix}
\omega^2_{0}  - \omega^2- i \gamma_1 \omega  & -i\omega\widetilde{\kappa}  \\
i\omega\widetilde{\kappa} & (\omega_{0}+\delta)^2 - \omega^2 - i \gamma_2 \omega 
\end{bmatrix}
\cdot
\begin{bmatrix}
x_{01}   \\
x_{02} 
\end{bmatrix}
 =
\begin{bmatrix}
gE{_0}   \\
{0} 
\end{bmatrix}
\end{equation} 
Here, $\omega_0=\omega_1=\omega_2-\delta$, $E_0$ is the external driving field amplitude, and $g$ is the coupling parameter between the bright mode and the far field. 
The coupling matrix models the amplitude of a single nanostructure, but to calculate the reflection and transmission amplitudes of scattered radiation from a metasurface array, we consider the average collective far-field emission of multiple antennas.\cite{deAbajo_MSscaling,Koenderink2020_MStransferMatrix}
\begin{equation}
r = \frac{2\pi ik}{\mathbb{A}}\frac{1}{\frac{1}{x_{01}} -\frac{2\pi ik}{\mathbb{A}}}   \label{EQ_main_deAbajoMSscaling_r}
\end{equation}
Here, $\mathbb{A}$ is the array's unit cell area, $k=\omega/c$ is the wavevector and the transmission amplitude is calculated as $t=1+r$.
The theoretical transmission intensity ($T=|t|^2$) of metasurface arrays at various coupling strengths are shown in Fig.~\ref{fig1_CFWintro}(c).

The real eigenvalues of an optical system correspond to radiative spectral modes. A dressed Hamiltonian for the system is derived from the coupled equations in Sec. SIII.II of the SI to reach an eigenvalue equation.
\begin{equation} \label{Eq_Eigenvalues}
\omega_{\pm} = \omega_0 - i \chi \pm \sqrt{\kappa^2 - \beta^2} 
\end{equation} 
$\chi = (\gamma_1 + \gamma_2)/2$ is the average loss rate of the constituent antennas, while $\beta = (\gamma_1 - \gamma_2)/2$ is their dephasing contrast. Weak coupling occurs for $\kappa<\beta$, while the strong coupling regime requires $\kappa>\beta$, and $\kappa=\beta$ is the exceptional point. 
As the coupling increases, the radiative resonance remains constant at $\mathrm{\omega_0}$, before bifurcating into positive and negative branches when the modes hybridise. This is seen in the transmission modes in Fig.~\ref{fig1_CFWintro}(c). 
The imaginary eigenvalues behave in a reverse manner, coalescing at the exceptional point to remain constant in the strong coupling regime.
Figure~S7 in Sec.~SIII.II of the SI shows the real and imaginary eigenvalues as function of the coupling.

The theory so far has assumed an external driving field composed of a single, purely real frequency value. CFW analysis instead proposes driving the system with a complex wave, $\widetilde{\omega} = \omega-i\tau/2$. Substitution into Eqn.~\ref{EQ_matrix_CoupledHarm} leads to the coupled mode equation,
\begin{equation}  
\label{EQ_matrix_CoupledHarm_CFW}
\begin{bmatrix}
\omega^2_{0}-\omega^2 - \Delta(\tau) - i\omega(\gamma_1-\frac{\tau}{2})  & -i\omega\widetilde{\kappa}-\widetilde{\kappa}\frac{\tau}{2}  \\
i\omega\widetilde{\kappa}+\widetilde{\kappa}\frac{\tau}{2} & (\omega_{0}+\delta)^2 -\omega^2-\Delta(\tau)-i\omega(\gamma_2-\frac{\tau}{2})
\end{bmatrix}
\cdot
\begin{bmatrix}
x_{01}   \\
x_{02} 
\end{bmatrix}
 =
\begin{bmatrix}
gE{_0}   \\
{0} 
\end{bmatrix} \mathrm{.}
\end{equation} 
 
A comparison of the two matrix equations shows that the imaginary component of the driving frequency reduces both loss rates, $\gamma_{1,2}\rightarrow \gamma_{1,2}-\tau/2$, and hence $\tau$ may be interpreted as a virtual gain. 
Even a narrow range, $0<\tau/\omega_0<0.1$ significantly improves the Fano interference visibility of the metasurface around the exceptional point, as seen in Fig.~\ref{fig1_CFWintro}(d). 
With the application of virtual gain, we inevitably induce dispersion through the additional term, $\Delta(\tau)=\frac{\tau\gamma}{2}-\frac{\tau^2}{4}$, which shifts the antennas' resonances. Even for typical plasmonic loss, $\gamma_i/\omega_0 \ll 1$ this shift is small when $\tau/\omega_0 \ll 1$, being essentially a second order correction. Finally, consider the effect of CFW on the coupling: from Eqn~\ref{EQ_matrix_CoupledHarm_CFW}, the coupling is modified in phase and amplidutde by $\widetilde{\kappa}\tau/2$, which for reasonable values of $\widetilde{\kappa}/\omega_0 \ll 1$ is also a second order correction. Furthermore, when using the rotating wave approximation, in the case of both small coupling and dephasing, the off-diagonal terms would be completely unaffected by the CFW analysis (see SI Section SIII.II).  Since the coupling strength remains invariant under CFW illumination to first order, we would expect the exceptional point to also remain constant, $
2\kappa_{EP} = (\gamma_1-\tau)-(\gamma_2-\tau) = \gamma_1-\gamma_2=\beta$. 

A propagating wave in free-space with a complex frequency $\widetilde{\omega}$ is challenging to produce experimentally. Fortunately, synthetic CFWs can be used to retroactively ``illuminate'' a system to similar effect. 
Here, we follow the methodology in Ref.\parencite{Zeng2024_CFW_MoleSens}, where a synthetic CFW is defined as $E(t_0) = E_0e^{-i\widetilde{\omega}t_0}\theta(t_0)$ and $\theta(t_0)$ is a truncation in time of $\theta(t_0)=0$ for $t_0<0$ and $\theta(t_0)=1$ for $t_0\geq0$. 
This truncation is necessary to prevent unphysical divergence at negative times and results in the waveform in Fig.~\ref{fig1_CFWintro}(e). 
Using its Fourier coefficients, the CFW is applied as a frequency domain convolution to a generalised sum over discrete scattering amplitudes, $P(\omega_n)$, which are related to the measured reflection and transmission spectra.
\begin{equation}
    P(\widetilde{\omega}) \approx \frac{1}{2\pi} \sum_{n} \widetilde{P}(\omega_n)\frac{1}{i(\widetilde{\omega}-\omega_n)} e^{i(\widetilde{\omega}-\omega_n)t_0} \Delta\omega
    \label{EQ_CFW_response_FT_discrete}
\end{equation}
This is a coherent summation across real frequency points and requires amplitude and phase information, $\widetilde{P}(\omega_n)=P(\omega_n)e^{i\phi(\omega_n)}$. 
Phase information need not be collected from the reflection and transmission spectra, as it can be calculated using the Kramers-Kronig relation. 
(See SI Sec.~SIV for the derivation of Eqn.~\ref{EQ_CFW_response_FT_discrete} and practical notes on the implementation of CFW analysis.)

\subsection{Nanofabrication and spectral response}
\begin{figure*}[ht!]
    \centering
    \includegraphics[width=0.99\linewidth]{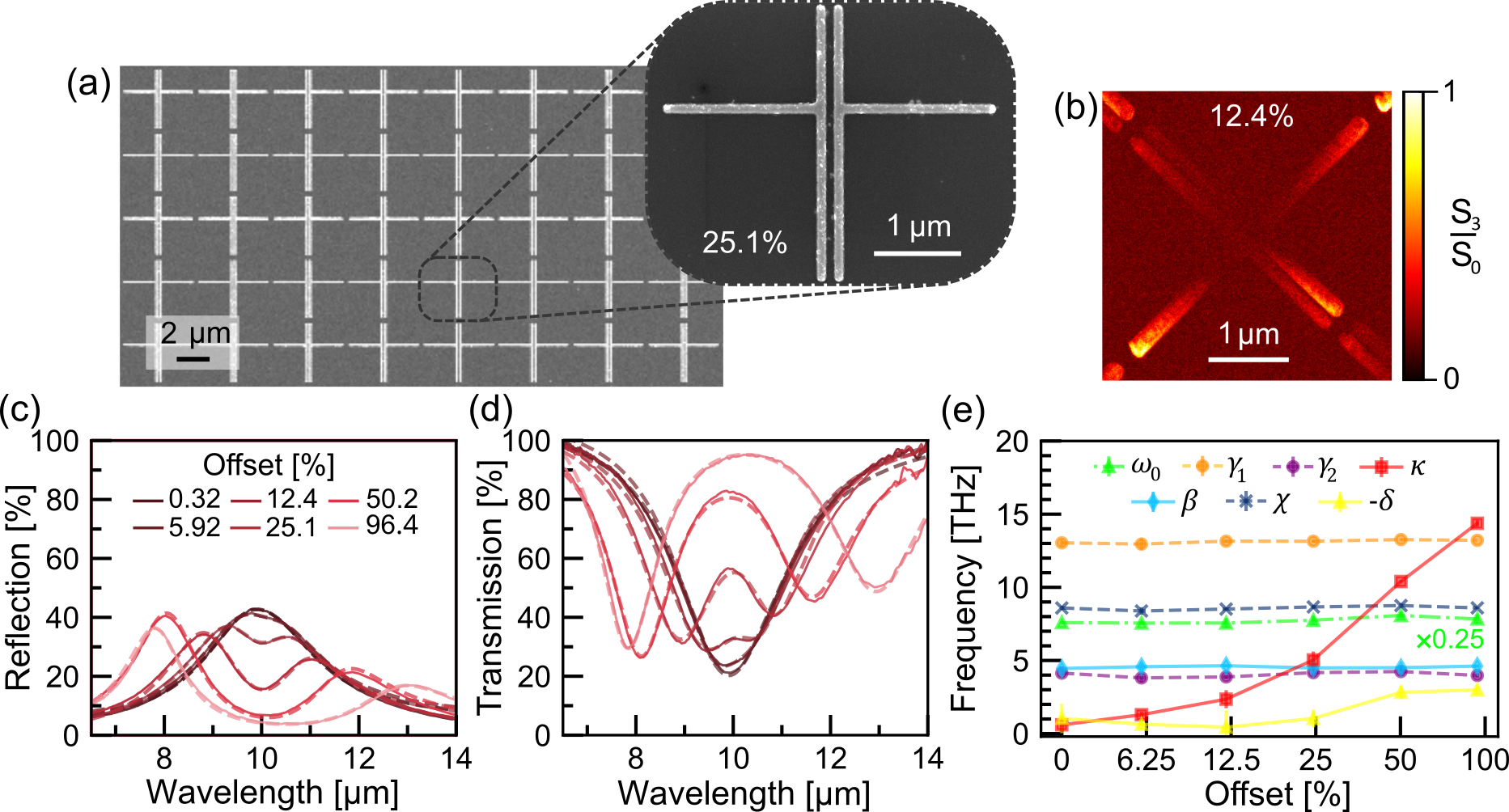}
    \caption{\textbf{Optical response and analysis of the plasmonic metasurface arrays}. (a) A section of a 100x100 \textmu m$^2$ metasurface array of the $25.1\%$ offset nanostructures with a single structure inset. (b) Near-field amplitude map of a  $12.4\%$ offset nanostructure excited on resonance. The scattered near-field light is demodulated at the third harmonic of the probe tapping frequency and normalised to the maximum signal. (c) Reflection and (d) transmission intensity of all metasurface arrays, where solid lines are experimental data and dashed lines are fits to the coupled-mode theory. (e) The parameters extracted by the coupled-mode theory at each offset. }
   
    \label{fig2_plas_nanostruc}
\end{figure*}

Composite plasmonic antennas were fabricated for a range of offset positions to control the inter-mode coupling strength. The gold nanostructures are $50$ nm in height and are bound to the \ch{CaF2} substrate by a $1.5$ nm chromium adhesion layer.
The individual nanostructures were fabricated into $100\times100$ \textmu m$^2$ arrays for large area illumination, an example of which is shown in Fig.~\ref{fig2_plas_nanostruc}(a).
The offset positions are measured from the SEM images of each nanostructure, as shown in the inset to Fig.~\ref{fig2_plas_nanostruc}(a), and compared to the designed offsets in Table~\ref{TABLE_BO_measured}. 
Fig.~S3 of the SI contains SEM images of the remaining nanostructures and Table SI compares their other measured dimensions to the designed dimensions, to validate the accuracy of the fabrication.  
\begin{table}[ht!]
\begin{center}
\begin{tabular}{ c |c c c c c c| }
 & \multicolumn{6}{c|}{Offset [\%]} \\
 \hline
 Design  & 0.00 & 6.25 & 12.5 & 25.0 & 50.0 & 97.0\\
 Fabrication & 0.32 $\pm$ 0.34 & 5.92 $\pm$ 0.48 & 12.4 $\pm$ 0.2 & 25.1$\pm$ 0.6 &  50.2 $\pm$ 0.4 & 96.4 $\pm$ 0.2  
\end{tabular}
\end{center}
\caption{\textbf{Measured offsets of the nanostructures compared to design}. The maximum offset is 97.0 to account for the finite width of the dimer antenna.
}
\label{TABLE_BO_measured}
\end{table}
The coupled modes of the system are further verified through scattering scanning near-field optical microscopy (sSNOM).
The E-field amplitude map in Fig.~\ref{fig2_plas_nanostruc}(b) is of a $12.4\%$ offset nanostructure excited on resonance and shows bright regions at either end of the dimer antenna and at either end of the waveguide.
This is characteristic of the two constituent resonators driven at their fundamental resonances\cite{Neuman2015_SNOM_PlasAntenna} and is in agreement with the near-field simulations in Fig.~S6.
It should be noted that sSNOM measurements are preferentially sensitive to the out-of-plane field component, and therefore, the amplitude within the waveguide is not observed to be enhanced.
A more detailed investigation of the near-field modes at a range of coupling strengths, conducted on similar, smaller nanostructures, can be found in Sec.~SII of the SI.
From this investigation, we are confident that the coupling mechanism is focussing the EM field into the waveguide gap.\cite{Xiao2024_nanofocus}
Details of the sSNOM system and measurements can be found the methods section.

The spectral responses of the metasurface arrays are measured using a Fourier transform infrared (FTIR) spectrometer in reflection and transmission. 
The theoretical transmission and reflection amplitudes are fitted to the intensity measurements in Fig.~\ref{fig2_plas_nanostruc}(c,d) to extract the parameters in Fig.~\ref{fig2_plas_nanostruc}(e). The main spectral feature due to extinction of a single mode occurs at $\sim10$ \textmu m for the symmetric antenna design. 
With increasing antenna offset, the modes eventually bifurcate around $12.4\%$. These modes further separate as offset and coupling increase; a feature of Rabi splitting that suggests the strong coupling regime has been reached.  
The coupled-mode theory (dashed lines) closely models the experimental measurements (solid lines), and the extracted coupling increases with offset as expected. 
The intersection of $\kappa$ and $\beta$ informs us that the system is in the strong coupling regime for offsets larger than 25.1\%. 
The apparent disagreement between the observed bifurcation and the extracted coupling is attributed to the contrast in loss between $\gamma_1$ and $\gamma_2$, leading to electromagnetically induced transparency effects (or Fano interference) in the weak coupling regime.\cite{Limonov2017_FanoResReview}  
Such a loss contrast is expected between a nonradiative plasmonic mode with loss limited by Drude damping and a bright mode with additional radiative damping.\cite{Liu2009_EIT?_plasDerudeLim}
The detuning parameter, $\delta$, is found to be small and increases slightly at larger offsets.
The detuning is also found to be negative across all offsets, meaning that the dimer is blue-shifted from the waveguide.  
The average resonance across the metasurface arrays is extracted at $\omega_0=9.71 \pm0.23$ \textmu m.

\subsection{CFW analysis assessment}
\begin{figure*}[ht!]
    \centering
    \includegraphics[width=0.985\linewidth]{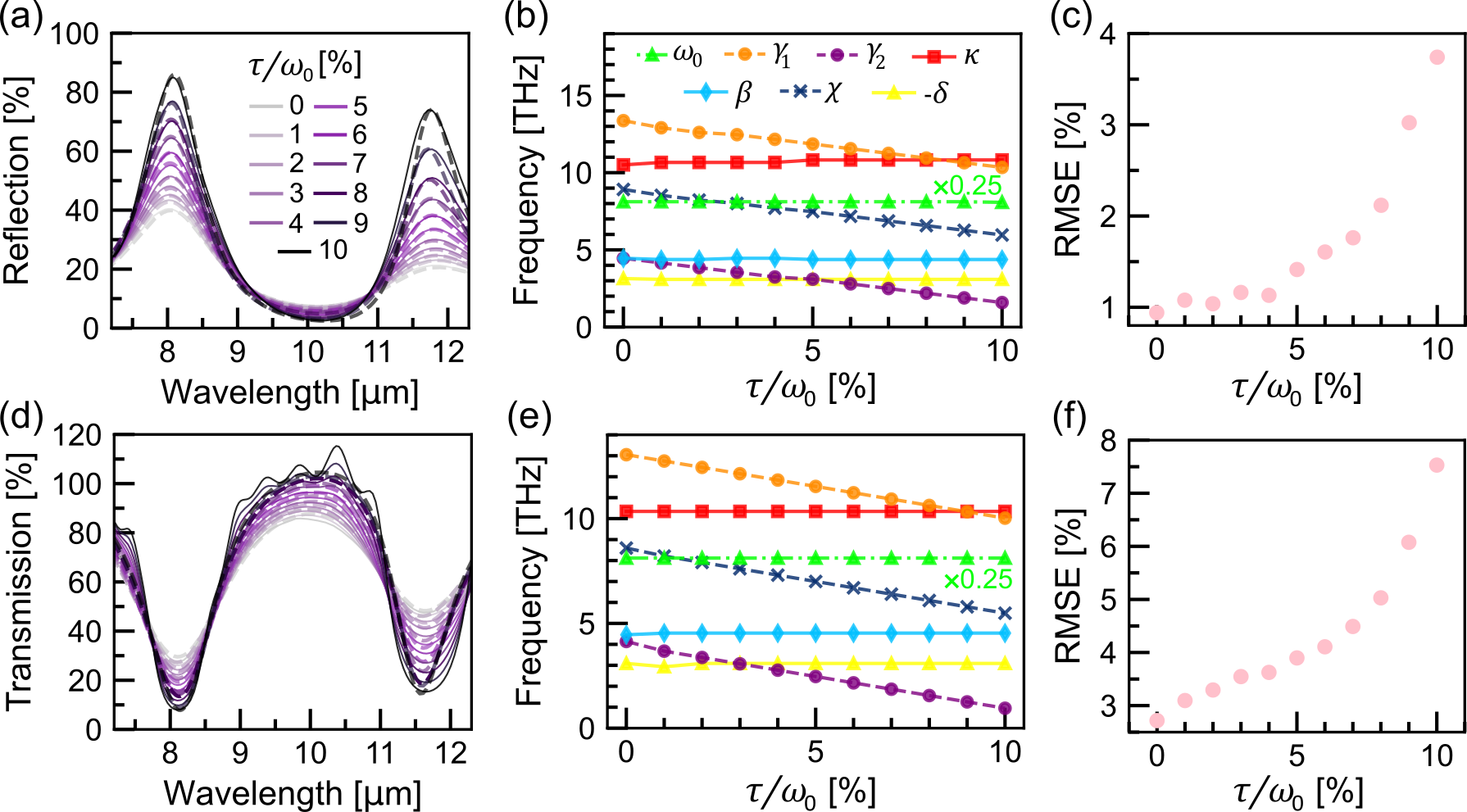}
    \caption{\textbf{Spectral response of plasmonic dimer-coupled resonant metasurface arrays at $\mathbf{50.2\%}$ with increasing virtual gain applied through CFW analysis}. (a) Reflection at various virtual gain values. Solid lines represent experimental data and dashed lines are fits to the coupled-mode theory. (b) Parameters extracted from the fits. (c) RMSE of the fits. (d-f) The corresponding analysis for transmission.  }
    \label{fig3_CFW_assessment}
\end{figure*}

In order to demonstrate that the coupling regime is invariant to the degree of loss compensation, progressively larger virtual gains are applied to the spectral intensity measurements of the $50.2\%$ offset metasurface in Fig.~\ref{fig3_CFW_assessment}. 
The coupled-mode theory is used to monitor changes in the resonance, loss, and coupling parameters.

The virtual gain is applied up to $\tau/\omega_0=10\%$, and as $\gamma_2/\omega_0\approx14\%$, over two thirds of the waveguide loss is compensated.
As the virtual gain increases, the spectral features narrow in Fig.~\ref{fig3_CFW_assessment}(a) and (d). 
The unphysical sidebands that appear at higher virtual gains and high transmissions are attributed to the temporal truncation of the complex wave, as discussed in Ref.\citenum{Zeng2024_CFW_MoleSens} and in Sec. SIV of the SI.
Although the amplitude of these features are reduced by averaging over time, they cannot be completely removed.

The extracted parameters in Fig.~\ref{fig3_CFW_assessment}(b) and (e) show the losses, $\gamma_{1,2}$ and their average $\chi$, falling linearly with increasing virtual gain.
As predicted, the coupling, $\kappa$, and the position of the exceptional point, $\beta$, remain essentially constant across the range of applied virtual gain. 
We also note that the resonant frequency, $\omega_0$, remains constant, confirming that the second order CFW induced shift in the resonance, $\Delta(\tau)$, is too small to be measured within the virtual gain range. 
Because both the dimer and waveguide resonances are subject to this shift, the detuning, $\delta$, is also unaffected by the virtual gain.

The root mean square error (RMSE) assesses how accurately the discrete dipole theory models an experimental result. The low RMSE for the original spectra in Fig.~\ref{fig3_CFW_assessment}(c) and (f), at $\tau/\omega_0=0\%$, instils confidence in the analytical model. The increase in RMSE with virtual gain, for both reflection and transmission, suggests that the CFW analysis becomes less physical; however, the fit to theory still remains satisfactory over the range considered. 
We attribute the growth in RMSE to the temporal truncation of the synthetic CFW, which could promote instability even before unphysical sidebands appear. 
Assessing the growth rate of the RMSE allows an upper limit of virtual gain to be conservatively established at $\tau/\omega_0=6\%$. This is approximately half of the loss of the dark waveguide resonance.
During the preparation of this manuscript, an improved algorithm to process the computation was demonstrated to significantly reduce such problems.\cite{Guan2026_HighOrderCFW_NatPhys}

The CFW analysis of reflection spectra appears to be more robust to the synthetic CFW illumination, with a lower RMSE recorded across the virtual gain range. This is attributed to the fact that, away from resonant features, transmission spectra approach 100\% at larger virtual gain and, therefore, the error introduced by the synthetic CFW's temporal truncation is more significant. 
In contrast, even at $\tau/\omega_0=10\%$, reflection never exceeds 90\%.

\subsection{Molecular coupling}
\begin{figure*}[ht!]
    \centering
    \includegraphics[width=0.999\linewidth]{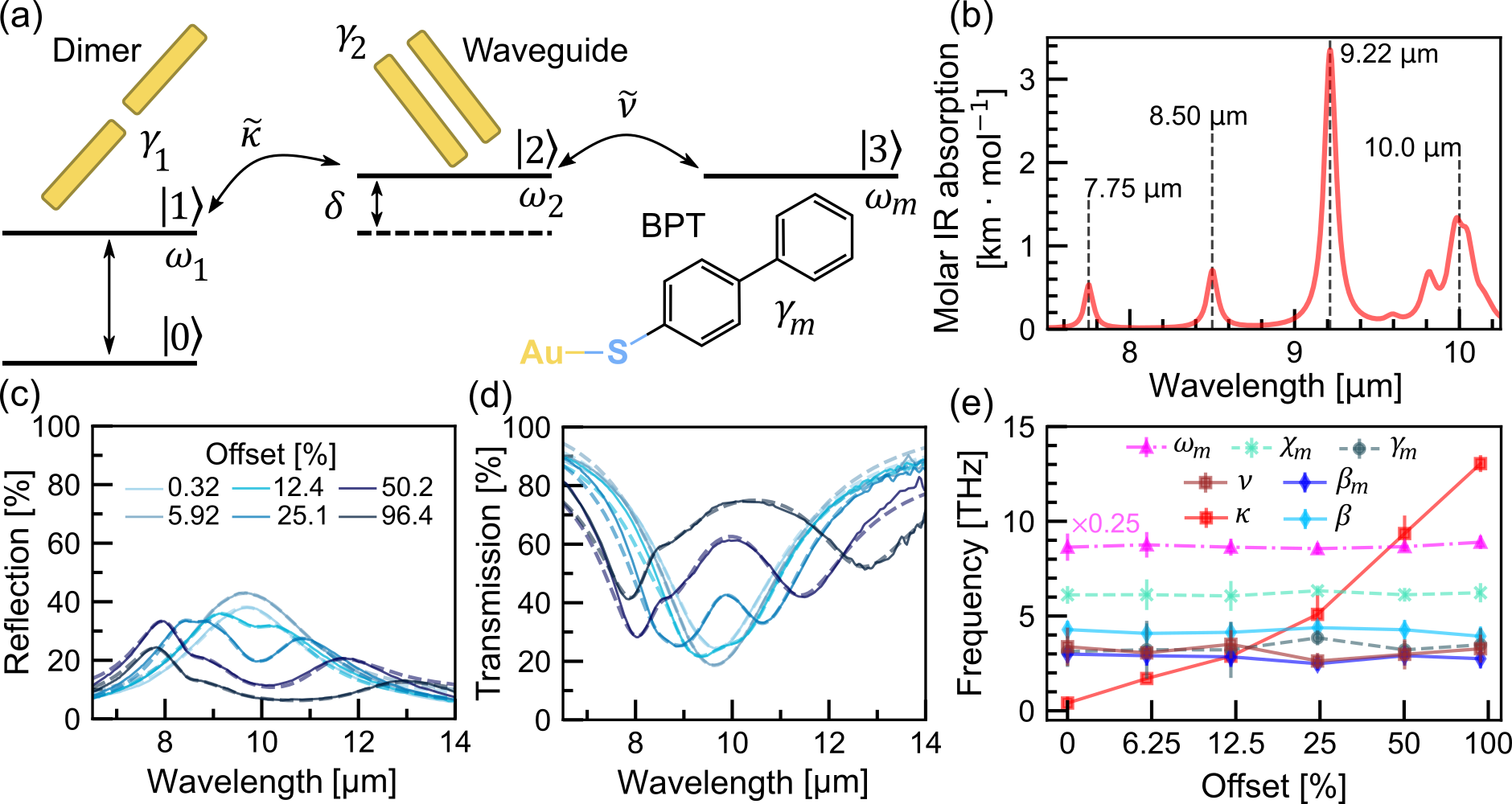}
    \caption{\textbf{Functionalisation of the plasmonic metasurfaces with BPT}. (a) Energy level diagram visualising the coupling in the nanostructure to excite the BPT molecules. (b) Molecular IR absorption of BPT, 
    taken from the Molecular Vibration Explorer Database.\cite{Benda2022_MoleVibrationDB} (c,d) Reflection and transmission of the metasurface arrays after coating with BPT. Solid lines are experimental measurements and dashed lines are fits to the coupled-mode theory. (e) Extracted molecular resonance and coupling parameters alongside the corresponding plasmonic parameters $\kappa$ and $\beta$.   }
   
    \label{fig4_BPTfuncationalisation}
\end{figure*}

The plasmonic metasurfaces are functionalised with a self-assembled monolayer (SAM) of biphenyl-4-thiol (BPT). The BPT molecule is composed of two benzene rings with a single thiol group (-SH) branching from one of the rings.
When in an ethanol solution and left in contact with a gold surface, the thiol ligand forms a thiolate-gold bond (S-Au), as shown inset to Fig.~\ref{fig4_BPTfuncationalisation}(a).\cite{Gronbeck2000_thiolateBond} 
Over time, many molecules adsorb to the surface to form a SAM.\cite{Ulam1996_SAMs}
When bound to the exposed gold of the metasurfaces, the plasmonic modes couple to the IR vibrational modes of the molecule.
Furthermore, the enhanced EM field associated with the waveguide gap is expected to result in stronger coupling to the molecular modes through the waveguide than through the dimer.
Therefore, the molecular modes are excited, as shown in the three-resonator energy level diagram in Fig.~\ref{fig4_BPTfuncationalisation}(a), $\ket{0} \rightarrow\ket{1} \rightarrow\ket{2} \rightarrow\ket{3}$, and it is assumed that no coupling occurs from free-space to the molecule, $\ket{0}\nrightarrow\ket{3}$, or through the dimer mode, $\ket{1}\nrightarrow\ket{3}$.
From the derivation in Sec.~SIII.III of the SI, the coupling matrix for the three-resonator system is defined.
\begin{equation}
\begin{bmatrix}
  \omega^{2}_0 - i\omega\gamma_1 - \omega^{2} & -i\omega\widetilde{\kappa} & 0 \\ 
  i\omega\widetilde{\kappa} & (\omega_0 + \delta)^{2} - i\omega\gamma_2 - \omega^{2} & -i\omega\widetilde{\nu} \\
  0 & i\omega\widetilde{\nu} & \omega^{2}_m - i\omega\gamma_m - \omega^2
\end{bmatrix}
\cdot
\begin{bmatrix}
  x_{01}  \\
  x_{02} \\
  x_{0m} \\
\end{bmatrix}
=
\begin{bmatrix}
  gE_0  \\
  0 \\
  0 \\
\end{bmatrix} \label{CM_MolecularCouplingMatrix}
\end{equation} 
The complex coupling between the waveguide and molecules is $\widetilde{\nu}=\nu e^{i\phi_\nu}$, where $\nu$ is the molecular coupling amplitude and $\phi_\nu$ is the phase.  The molecular resonance and corresponding loss are denoted by $\omega_m$ and $\gamma_m$ respectively. We note that this may be generalised for any number of vibrational modes of the system.
 
Fig.~\ref{fig4_BPTfuncationalisation}(b) shows the IR vibrational spectra of a BPT molecule with a thiolate-gold bond across the relevant spectral range, taken from the Molecular Vibration Explorer Database.\cite{Benda2022_MoleVibrationDB} After the metasurfaces are functionalised with BPT, the spectral intensities in Fig.~\ref{fig4_BPTfuncationalisation}(c) and (d) show a single Fano-interference feature $\sim8.50$ \textmu m, at offsets $>12.4\%$.
The three-resonator model fits the experimental data well to extract the molecular resonance parameters in Fig.~\ref{fig4_BPTfuncationalisation}(e).
The plasmonic coupling parameters found in Fig.~\ref{fig2_plas_nanostruc}(e) are used as bounds for the three-resonator system, and $\kappa$ and $\beta$ are shown in Fig.~\ref{fig4_BPTfuncationalisation}(e).
Fig. S9 in SI Sec.~SV compares the parameters of the plasmonic resonators extracted using the two- and three-resonator models. 
The molecular resonance is measured at $8.64\pm0.17$ \textmu m and is identified as the 8.50 \textmu m resonance in Fig.~\ref{fig4_BPTfuncationalisation}(b).
The molecular coupling amplitude remains constant at $\nu=3.15\pm0.47$ THz regardless of $\kappa$, since coupling is only varied between the plasmonic resonators. Molecular loss, $\gamma_m$, is 3.36$\pm$0.36 THz, only slightly less than $\gamma_2$ at 4.08$\pm$0.30 THz and similar in magnitude to $\nu$.

As $\kappa$ increases, the metasurfaces transition through the exceptional point, and $\omega_+$ spectrally tunes to the 8.50 \textmu m resonance. 
From this description, an eigenvalue equation to be written for the three-resonator system in a similar format to Eqn.~\ref{Eq_Eigenvalues}.
\begin{equation}
\omega_{m\pm} = \omega_m - i\chi_m \pm \sqrt{\nu^2 - \beta_m^2 }\label{main_CM_Eq_moleculeEigenvalues}
\end{equation}  
Here, degeneracy is necessarily assumed between the hybrid plasmonic and molecular modes, $\omega_{+} =\omega_{m}$. 
As the molecules are coupled through a hybridised plasmonic mode, the combined dimer-waveguide loss parameter $\chi$ must be considered. 
Like before, we can define an exceptional point at $\nu=\beta_{m}=(\chi-\gamma_m)/2$, and an average loss of $\chi_m = (\chi+\gamma_m)/2$. 
These parameters are also visualised against offset in Fig.~\ref{fig4_BPTfuncationalisation}(e). 
We find $\beta_m=2.81 \pm 0.50$ and so $\nu>\beta_m$.
However, within the uncertainties, both the weak and strong regimes can be reached, along with the plasmonic-molecular exceptional point. Therefore, a definitive claim of strong coupling to the molecule cannot be made. 

\subsection{Uncovering IR molecular modes}
\begin{figure*}[ht!]
    \centering
    \includegraphics[width=0.999\linewidth]{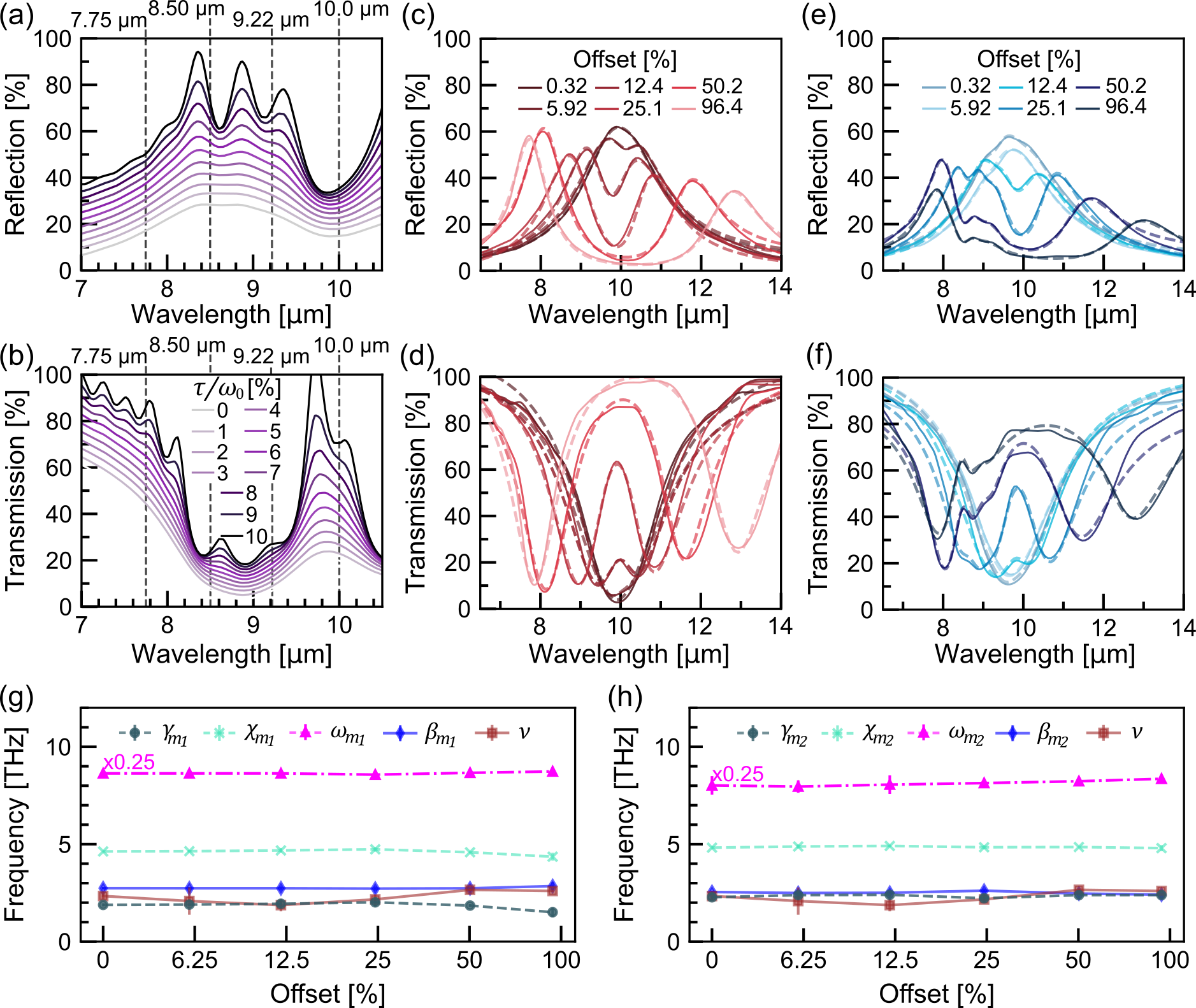}
    \caption{\textbf{CFW analysis on BPT coated plasmonic metasurfaces}. 
    (a) Reflection and (b) transmission intensity of the 25.1\% offset metasurface, across the spectral range where BPT resonances are expected, at various virtual gain values. 
    Dashed vertical lines mark BPT’s MIR absorption modes.
    (c) Reflection and (d) transmission of the uncoated metasurfaces with a virtual gain of $\tau/\omega_0 = 6\%$ applied. 
    (e,f) The equivalent spectral intensities after coating with BPT at the same virtual gain. 
    Solid lines represent experimental measurements and dashed lines are fits to the coupled-mode theory.
    Extracted coupling and resonance parameters for the (g) higher and (h) lower energy resonances uncovered by CFW analysis.  }
   
    \label{fig5_UncoveringIRmodes}
\end{figure*}

The same range of virtual gain, up to $\tau/\omega_0=10\%$, is applied to the spectral response of the metasurfaces coated with BPT.
The resulting spectral intensities are shown for the $25.1\%$ offset metasurface in Fig.~\ref{fig5_UncoveringIRmodes}(a) and (b), across a spectral range that covers the IR BPT modes of interest.
The application of synthetic CFWs reduces the plasmonic and molecular losses, and the 8.50 \textmu m resonance becomes more pronounced.
Around 9.22 \textmu m, a mode appears at $\tau/\omega_0\ge5\%$ that was initially obfuscated by loss.
The mode is clearer in reflection but can also be seen in transmission.
The instability at larger $\tau/\omega_0$ is problematic when uncovering modes that are initially completely obscured. 
For example, at $\tau/\omega_0\ge8\%$ in transmission, the unphysical sidebands spectrally overlap with the molecular modes expected at 7.75 and 10.0 \textmu m.
In this case, it is useful to qualitatively analyse the processed BPT coated spectra in parallel with the equivalent uncoated spectra, under the assumption that the sidebands appear similar in spectral regions with comparable resonant features.
This qualitative analysis is conducted in Sec.~SVI of the SI, where it is concluded that while the 10.0 \textmu m molecule resonance is likely uncovered, it requires a high virtual gain that will make the system difficult to model.

In order to reduce the effect of the instability at high virtual gain, the upper limit of $\tau/\omega_0=6\%$, established above, is applied to all spectral data. The processed spectra in Fig.~\ref{fig5_UncoveringIRmodes}(c-f) show minimal evidence of sidebands, while the 8.50 and 9.22 \textmu m resonances are clearly present in the BPT coated spectra.

The coupled-mode theory is adapted to include two molecular resonators (a four-resonator model) where the respective molecular modes are labelled $\omega_{m_1,m_2}$ and $\gamma_{m_1,m_2}$. 
As before, the plasmonic parameters are extracted from the uncoated spectra using the two-resonator model (shown in Fig.~\ref{fig5_UncoveringIRmodes}(c) and (d)) to define bounds for the four-resonator model, shown in Fig.~\ref{fig5_UncoveringIRmodes}(e) and (f).
The four-resonator model now enables the parameters of the two molecular modes to be extracted simultaneously; shown in Fig.~\ref{fig5_UncoveringIRmodes}(g) and (h).

Fig.~\ref{fig5_UncoveringIRmodes}(g) displays the parameters of the higher energy molecular mode, where the resonant frequency, $\omega_{m_1}$, is averaged across all offsets to equal $8.67 \pm 0.08$ \textmu m and is attributed to the 8.50 \textmu m resonance of BPT.  It is therefore appropriate to compare these parameters to those of the original data in Fig.~\ref{fig4_BPTfuncationalisation}(e).  
Predictably, $\mathrm{\gamma_{m_1}}$ ($\mathrm{\gamma_{m}}$) has been artificially reduced by the virtual gain from $3.35 \pm 0.81$ THz to $1.85 \pm 0.25$ THz, and in turn, the average $\mathrm{\chi_{m_1}}$ ($\mathrm{\chi_{m}}$) has also fallen. 
However, it appears that the coupling amplitude $\mathrm{\nu}$ has reduced from $3.14 \pm 0.67$ to $2.29\pm 0.39$ THz, although this is within uncertainty bounds. 
This may not inherently be caused by the synthetic CFWs, and could be a result of a more accurate parameter extraction due to the reduction of loss. 
The Fano-type interference in Fig.~\ref{fig4_BPTfuncationalisation}(c) and (d) was attributed to one molecular resonance that has now been revealed to be two.  
This merging of two broader modes will skew the measured molecular parameters, including coupling. 
The uncertainty in these parameters has also decreased, which suggests we have a more accurate picture of the molecular coupling. 
As $\beta_{m_1}$ ($\beta_m$) remains unchanged at $2.75 \pm 0.06$, a consequence of $\nu$ falling is that the system is conclusively found to be in the weak coupling regime, $\nu<\beta_{m_1}$, even though the Fano interference effect is clearly pronounced with virtual gain. 

Fig.~\ref{fig5_UncoveringIRmodes}(h) shows the parameters of the initially hidden molecular mode. 
The extracted resonant frequency of $\omega_{m_2} = 9.23 \pm 0.22$ \textmu m confirms it as the $9.22$ \textmu m BPT mode. 
Intuitively, the reduced loss rate, $\mathrm{\gamma_{m_2}}$, is higher than $\mathrm{\gamma_{m_1}}$, at $2.34 \pm 0.09$ THz, causing $\beta_{m_2}$ to be $2.51 \pm 0.10$ THz and to overlap with $\nu$ within uncertainties. 
As a result, the mode sits around the exceptional point, between the weak and strong coupling regimes.

The application of CFWs to artificially reduce losses has led to a more accurate understanding of the $8.50$ \textmu m BPT resonance and the discovery and characterisation of the initially hidden $9.22$ \textmu m resonance.
Without the loss compensation, incorrect conclusions regarding the coupling strength between the plasmonic and molecular modes would have been reached.

\section*{Conclusions}

In this work, synthetic CFWs were applied to a coupled system of plasmonic and molecular resonators to enhance the spectral visibility of resonant modes by compensating for dephasing and loss. This approach enables the recovery of obscured molecular vibrational modes and allows their internal coupling energies to be quantified with confidence.

Within our analysis, we found that CFWs compensate for dephasing due to a virtual gain but leave the inter-mode couplings of a system invariant to first order. Here, we have used this feature to quantitatively assess opto-mechanical coupling between light in a highly confined plasmonic mode and the IR vibrational modes of a molecule. Although clear Fano interference was evident in the spectra, the CFW analysis showed that the opto-mechanical coupling was near the exceptional point that straddled the strong and weak coupling regimes. We also confirmed that the position of the exceptional point remained invariant under the virtual gain of CFWs. Importantly, we conclude that CFWs do not allow a weakly coupled system to become strongly coupled through virtual gain or vice-versa.

Collectively, these findings demonstrate the utility of synthetic CFW as a quantitative tool while also highlighting the practical limitations associated with high virtual gain. Given the prevalence of Ohmic losses in plasmonic systems, yet their excellent sensitivity to local environments, synthetic CFW represents a powerful post-processing strategy for enhancing plasmonic spectroscopic sensing.

\section*{Methods} \label{sec:methods}

\subsection*{Sample fabrication}
The metasurface arrays were fabricated by electron beam lithography on a 350 \textmu m thick \ch{CaF2} substrate. The substrate was coated with the positive tone resist polymethyl methacrylate (PMMA) and baked at 180 $\mathrm{^{o}}$C for at least 180 s. Nanostructure geometries are defined by electron beam exposure at 20 keV using a 10 \textmu m aperture (Raith E-line). The PMMA was developed in a 3:1 isopropyl alcohol (IPA) to methyl isobutyl ketone (MIBK) solution for 30 s, and submerged in IPA for a further 30 s to halt the development. A 1.5 nm adhesion layer of Cr was deposited by thermal evaporation onto the substrate, followed by 50 nm of Au (Angstrom Engineering Amod). PMMA and excess Cr/Au were removed by leaving the substrate in acetone for at least 24 hours and oxygen plasma ashing.

\subsection*{Optical measurements}
The spectral responses of the metasurface arrays were measured under an IR microscope (Bruker Hyperion 3000) coupled to a FTIR spectrometer (Bruker Vertex 70) through a $100\times100$ \textmu m square aperture. 
Reflection measurements were normalised to the intensity from a gold mirror and transmission measurements were normalised to a bare section of substrate. The IR intensities were recorded on a cryogenically cooled mercury cadmium telluride (MCT) detector. 

Near-field images were collected using a pseudoheterodyne detection capable sSNOM system (NeaSpec neaSNOM) in reflection with a 60$\mathrm{^o}$ zenith angle of illumination. 
P-polarised radiation, relative to the sample plane, from a tunable MIR QCL (Daylight Solutions MIRcat) was focused onto a given metasurface with the in-plane polarisation component aligned to the longitudinal axis of the dimer antennas. 
An uncoated Si tip with a 10 nm radius of curvature (Budget Sensors Tap300-G) was driven at 300 kHz, and the scattered signal, recorded by a cryogenically cooled MCT detector, was demodulated at the third harmonic of the driving frequency.

\section*{Data availability}
Data available upon request.

\section*{Supporting Information}
The supporting information contains numerical simulations and details on the fabrication of the plasmonic metasurfaces, near-field (sSNOM) measurements, coupled-mode theory derivations for each number of resonators, CFW theory, a comparison of coupling parameters extracted with two- and three-resonator models, and analysis of uncoated and BPT coated spectra at increasing virtual gain to identify uncovered molecular modes.

\section*{Acknowledgments}
T.S.M. acknowledges the support of the UK government Department for Science, Innovation and Technology through the UK National Quantum Technologies Programme and the National Physical Laboratory. This work is supported by the UK Engineering and Physical Sciences Research Council (EPSRC) through the Reactive Plasmonics EP/M013812/1 and Catalysis Plasmonics EP/W017075/1 Programme Grants.

\section*{Author Contributions}
T.S.M. designed and fabricated the metasurface arrays, collected far-field optical measurements and performed data analysis. 
X.X., R.F.O. and T.S.M. developed the coupled-mode theory. 
X.X. and R.F.O. created the code to implement the CFW analysis. 
T.S.M. wrote the paper with contributions from all co-authors. 
T.S.M., M.R.-G., T.A.-V. and N.J.H. collected near-field optical measurements. 
R.F.O., O.K., C.C.P and S.A.M. conceived and supervised the project.

\section*{Disclosures}
The authors declare no conflicts of interest.

\bibliography{CFW.bib}
\end{document}


\maketitle

\section{Simulation \& fabrication of nanostructures}
\begin{figure*}[ht!]
    \centering
    \includegraphics[width=0.70\linewidth]{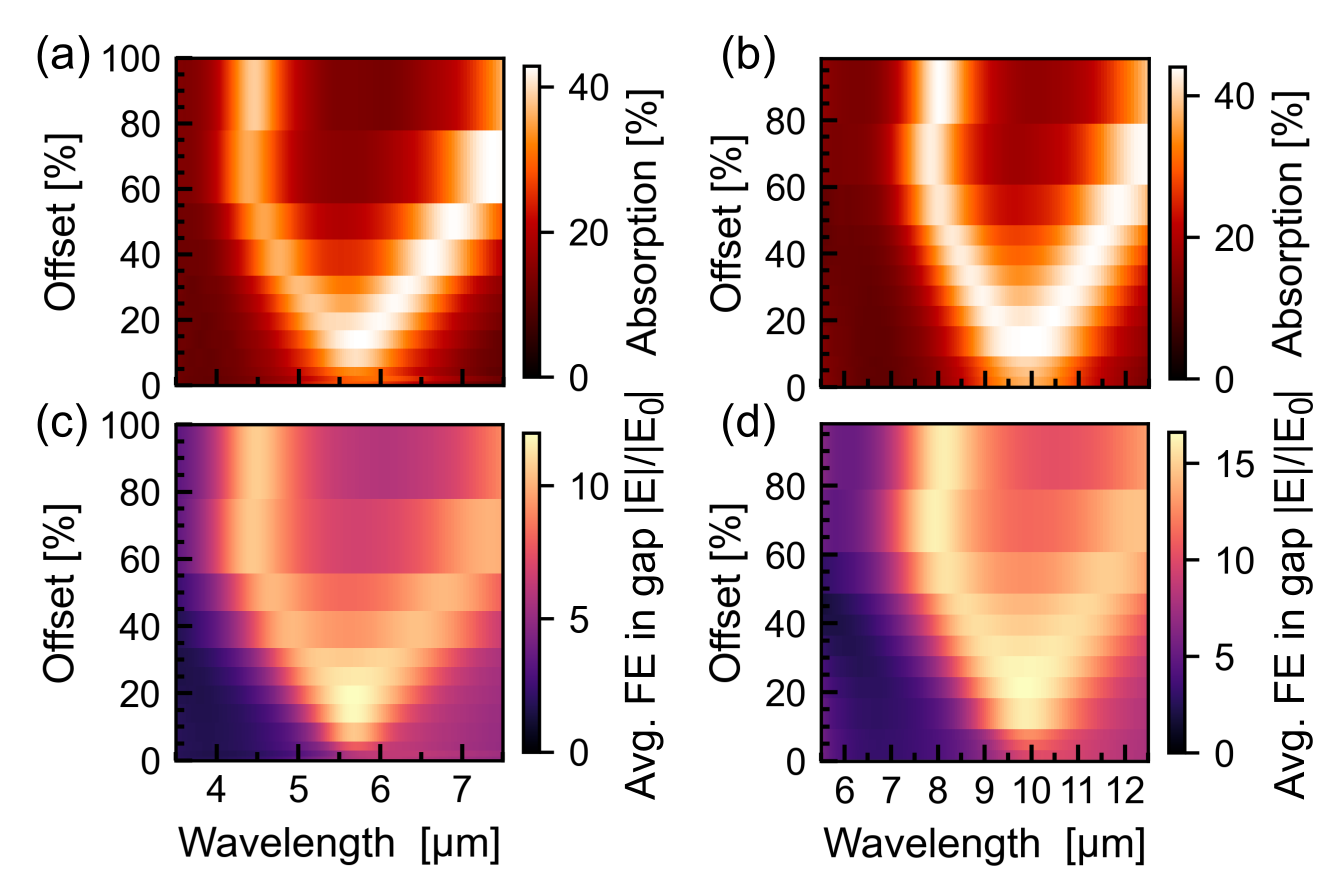}
    \caption{\textbf{Numerical simulations (FDTD) of the optical response of the coupled waveguide and dimer in the far and near field}. (a) Normalised absorption spectra of nanostructures resonant at 5.7 \textmu m ($L_{ant} = 1120$ nm, $L_{wg} = 1800$ nm, $W_{wg} = 100$ nm, $W_{au} = 100$ nm). (b) Normalised absorption spectra of nanostructures resonant at 10 \textmu m  ($L_{ant} = 1930$ nm, $L_{wg} = 3300$ nm, $W_{wg} = 100$ nm, $W_{au} = 100$ nm). (c,d) The normalised field enhancement (FE) in the waveguide gap, averaged over the in-plane area 5 nm above the gold. Maps that share an x-axis are for the same nanostructures. A 10 nm mesh unit cell has been used. } \label{Fig_SI_CutWG_FDTD}
\end{figure*}

Numerical simulations in the time domain (FDTD, Lumerical) are used to design the plasmonic dimer-coupled waveguides to reach degeneracy between the resonant components. 
The desired resonance of the waveguide was set by $L_{wg}$ and $W_{gap}$, and a range of $L_{ant}$ values were simulated to reach equal depths of the hybridised transmission modes and optimised field enhancement in the waveguide gap. 
The simulated spectral responses in the far and near field are shown in Fig.~\ref{Fig_SI_CutWG_FDTD}, as coupling is controlled by increasing the percentage offset of the dimer along the waveguide. 
The target degenerate resonances of 5.7 and 10 \textmu m were selected based on the spectral overlap with the IR modes of BPT and the peak power supplied by the quantum cascade laser (QCL) source used for subsequent near-field measurements.
The simulations apply normally incident plane wave excitation, polarised along the longitudinal dimer axis, and periodic boundary conditions to mimic arrays of repeated structures. 
In Fig.~\ref{Fig_SI_CutWG_FDTD}(a) and (b), the absorption increases with the offset at the designed wavelength before bifurcating into hybrid dimer and waveguide modes in the strong coupling regime.
The average field enhancement in the waveguide gaps, in Fig.~\ref{Fig_SI_CutWG_FDTD}(c) and (d), takes the ratio of the near-field amplitude to the incident field amplitude across the waveguide in-plane.
Here, the coupling into the waveguide increases at 5.7 and 10 \textmu m before splitting into two modes, slightly offset from the bifurcation observed in absorption.

The near-field coupling between compound nanostructures is also considered to avoid strong coupling and mode hybridisation.
Using simulations, the 0\% offset structures were iteratively separated, blue-shifting the hybrid dimer mode until it spectrally converged.\autocite{Nordlander2004_PlasDimerHybridTheory,Muskens2007_DimerResTuning} 
A $200$ nm spacing is deemed safely within the weak coupling regime for both resonant wavelengths. 

At 98.5\% dimer offset, field hotspots appear on the antenna corners that sit proud of the waveguide in-plane. 
While not inherently a negative outcome, it departs from the design of the coupled system. 
In fact, this additional field concentration red-shifts the split modes in Fig.~\ref{Fig_SI_CutWG_FDTD}.

\begin{figure*}[ht!]
    \centering
    \includegraphics[width=0.925\linewidth]{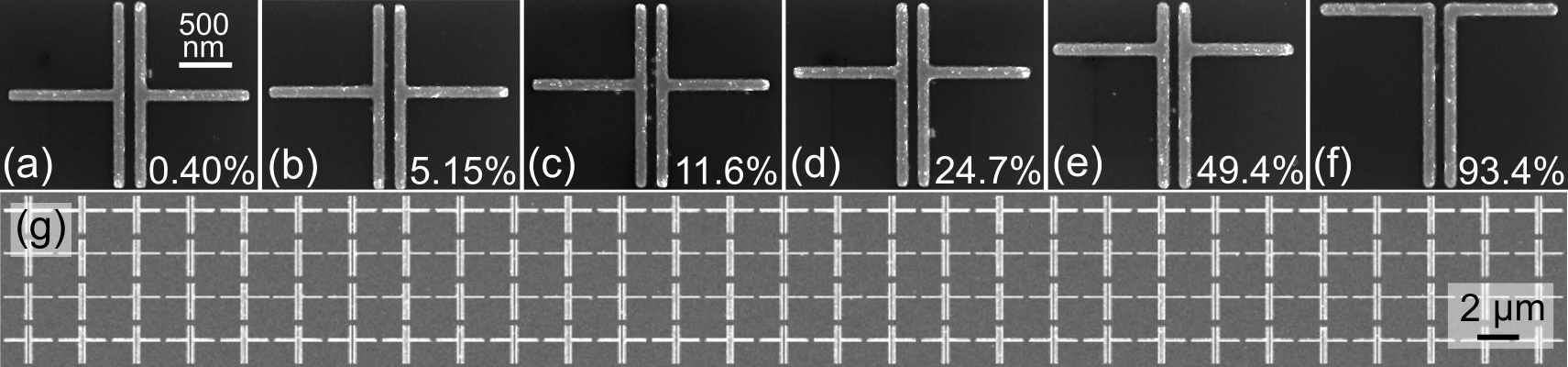}
    \caption{\textbf{SEM images of dimer coupled waveguide structures designed to support a 5.7 \textmu m resonance, fabricated via EBL}. 
    (a-f) Single nanostructures with their measured offset labelled. 
    (g) An area of a 100x100 \textmu m$^2$ metasurface array at $24.7\%$ offset.  }
   
    \label{fig_SI_CutWG_5-7SEM}
\end{figure*}

\begin{figure*}[ht!]
    \centering
    \includegraphics[width=0.925\linewidth]{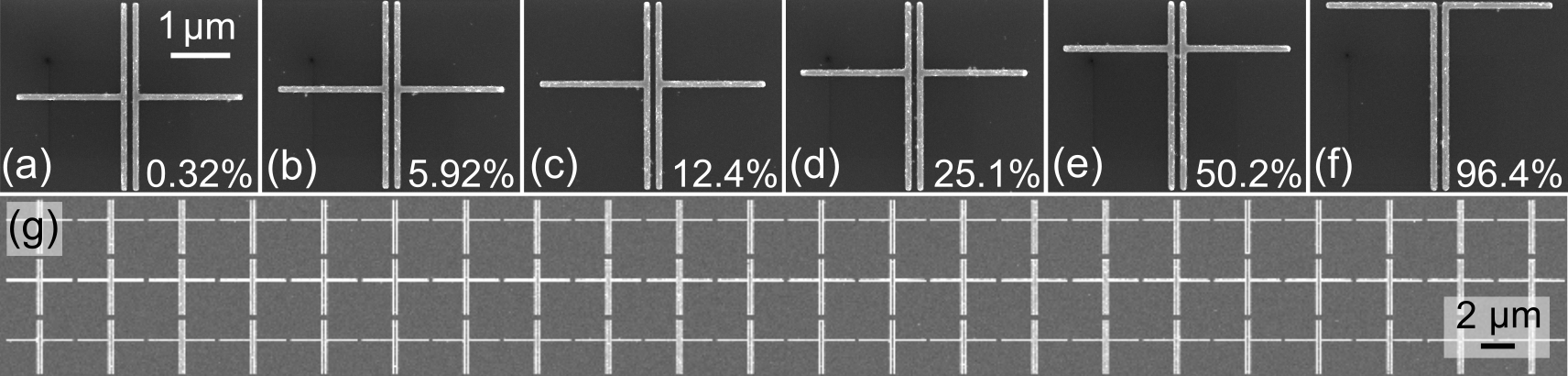}
    \caption{\textbf{SEM images of dimer-coupled waveguide nanostructures designed to support a 10 \textmu m resonance, fabricated via EBL}. 
    (a-f) Single structures with the measured offset labelled. 
    (g) An area of a 100x100 \textmu m$^2$ metasurface array at $25.1\%$ offset.}
   
    \label{Fig_CutWG_10SEM}
\end{figure*}

The simulated nanostructures of the dimensions noted in Fig.~\ref{Fig_SI_CutWG_FDTD} were fabricated into metasurface arrays on a \ch{CaF2} substrate using an electron beam lithography (EBL) system (eLine plus, Raith) at 20 kV. 
The gold structures are 50 nm in height with a 1.5 nm chromium adhesion layer.
The fabrication process is outlined in the methods section of the main text.
Scanning electron microscopy (SEM) images of single nanostructures resonant at 5.7 and 10 \textmu m wavelength, and at a range of offsets can be seen in Fig.~\ref{fig_SI_CutWG_5-7SEM}(a-f) and Fig.~\ref{Fig_CutWG_10SEM}(a-f), respectively.
The bright patches on the gold, mostly at the end of the resonators, are from residual PMMA that was not removed in the lift-off stage. 
After these images were taken, the samples were placed in a plasma asher (Diener Femto) for 3 minutes at 100 W power, followed by a 12 hr soak in acetone to remove the residual resist. 
Each single structure is repeated across an $100\times100$ \textmu m$^2$ array to form a metasurface.
An example of metasurface arrays at $24.7 \pm 0.5\%$ and $25.1\pm 0.6\%$ offset for the 5.7 and 10 \textmu m resonant wavelength structures can respectively be seen in panels (g) of  Fig.~\ref{fig_SI_CutWG_5-7SEM} and  Fig.~\ref{Fig_CutWG_10SEM}.
The nanostructures nominally resonant at 10 \textmu m are investigated in the main text and the nanostructures nominally resonant at 5.7 \textmu m are used to study the near field distribution of the dimer-coupled waveguide system.
The fabricated dimensions match the design for both sets of nanostructures within reasonable uncertainties, as shown in Table \ref{TABLE_dimension_measured} and \ref{TABLE_BO_measured}. 
The distance between nanostructures across all arrays is $\sim200$ nm, in agreement with simulations.

\begin{table}[ht!]
\begin{center}
\begin{tabular}{ c |c c c c| }
 & \multicolumn{4}{c|}{Dimension [nm]} \\
   Structure [\textmu m] & $L_{ant}$ & $L_{wg}$ & $W_{gap}$ & $W_{au}$ \\
 \hline
 5.7 design & 1120 & 1800 & 100 & 100 \\
 5.7 measured & 1117 $\pm$ 5 & 1797 $\pm$ 10 & 97 $\pm$ 7 & 103 $\pm$ 6 \\
 10 design & 1930 & 3300 & 100 & 100 \\ 
 10 measured & 1931 $\pm$ 5 & 3300 $\pm$ 11 & 101 $\pm$ 4 & 101 $\pm$ 1  
\end{tabular}
\end{center}
\caption{\textbf{Measured dimensions compared to design for both sets of nanostructures}.}
\label{TABLE_dimension_measured}
\end{table}

\begin{table}[ht!]
\begin{center}
\begin{tabular}{ c |c c c c c c| }
 & \multicolumn{6}{c|}{Offset [\%]} \\
 Structure [\textmu m]  & 0.00 & 6.25 & 12.5 & 25.0 & 50.0 & 94.0/97.0\\
 \hline
 5.7  & 0.40 $\pm$ 0.11 & 5.15 $\pm$ 0.39 & 11.6 $\pm$ 0.2 & 24.7 $\pm$ 0.5 & 49.4 $\pm$ 0.4 & 93.4 $\pm$ 0.7\\  
 10 & 0.32 $\pm$ 0.34 & 5.92 $\pm$ 0.48 & 12.4 $\pm$ 0.2 & 25.1$\pm$ 0.6 &  50.2 $\pm$ 0.4 & 96.4 $\pm$ 0.2  
\end{tabular}
\end{center}
\caption{\textbf{Measured offsets compared to design for both sets of nanostructures}. 
The largest designed offset is 94\% for 5.7 \textmu m and 97\% for 10 \textmu m}
\label{TABLE_BO_measured}
\end{table}

\section{Near-field measurements \& simulations} \label{SI_sec_sSNOM}
\begin{figure*}[ht!]
    \centering
    \includegraphics[width=0.999\linewidth]{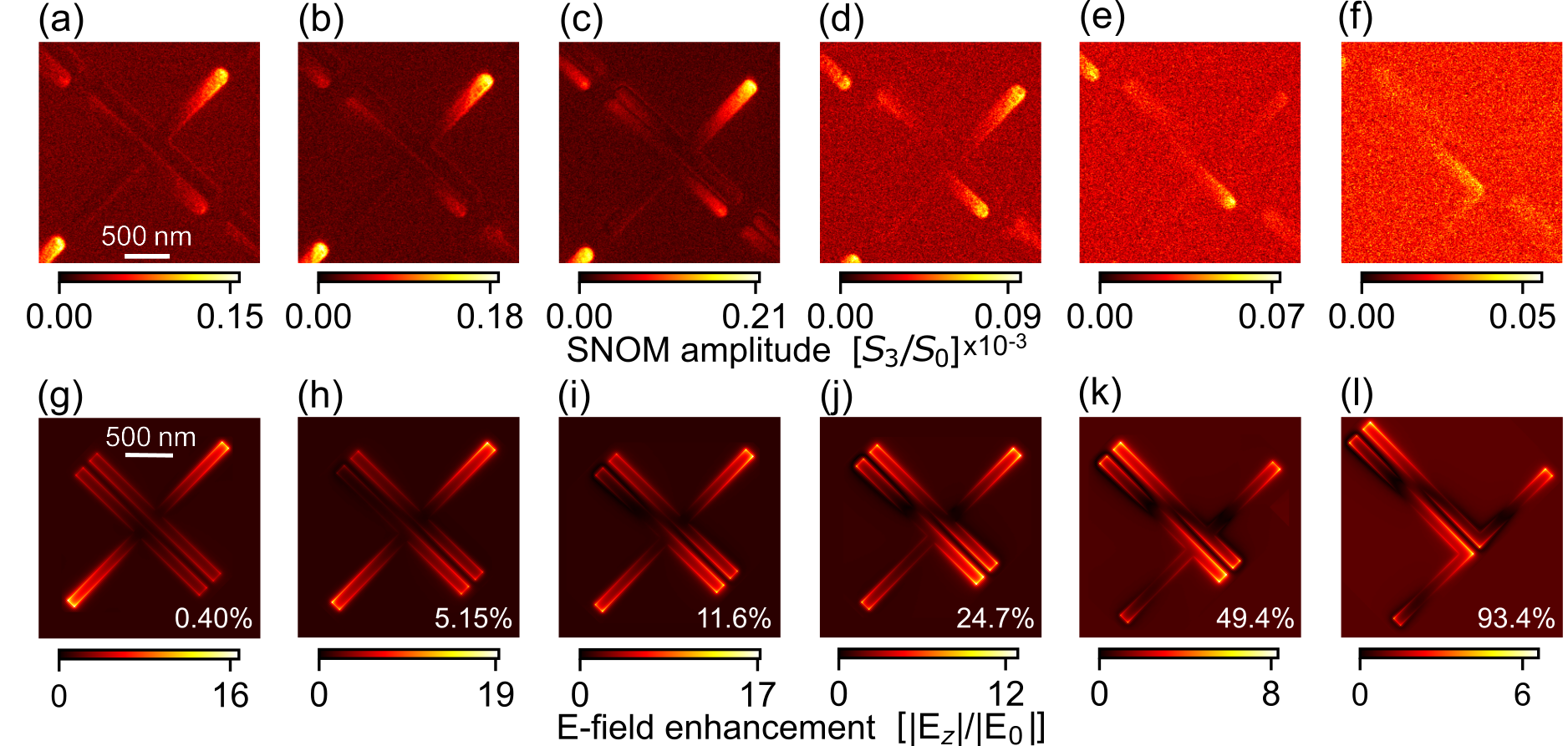}
    \caption{\textbf{sSNOM amplitude and numerical simulations (FDTD) of the resonant E-field enhancement across the 5.7 \textmu m nanostructures}. 
    (a-f) sSNOM amplitude, demodulated at the third harmonic and normalised to the raw signal on the MCT detector. 
    (g-l) FDTD simulations of the near-field amplitude, without accounting for tip interaction. 
    The percentage offset of the dimer along the waveguide is annotated on (g-l) for each column. }      
    \label{FIG_CutWG_SNOM_amp}
\end{figure*}

The near-field measurements, Fig.~\ref{FIG_CutWG_SNOM_amp}(a-f), were taken using a pseudoheterodyne detection capable sSNOM (neaSNOM, NeaSpec) system with p-polarised illumination and co-polarised collection relative to the sample plane.\autocite{Ocelic2006_Pseudoheterodyne} 
The in-plane component of the polarisation was aligned to the dimer's longitudinal axis. 
An uncoated Si tip with a 10 nm radius of curvature (Budget Sensors Tap300-G) was driven at 300 kHz, and the scattered signal was demodulated at the third harmonic of the driving frequency. 
The collected near-field amplitude is normalised to the raw mercury cadmium telluride (MCT) detector signal to account for power fluctuations. 
For the numerical simulations in Fig.~\ref{FIG_CutWG_SNOM_amp}(g-l), a p-polarised plane wave was angled at 60$^{\mathrm{o}}$ from the normal and the polarisation was aligned to the dimer's longitudinal axis to match the experimental excitation. 
An E-field monitor 5 nm above the gold surface calculated only the E$_z$ component to mimic the polarisation selectivity of the sSNOM tip.
The perturbation of the near field by the sSNOM tip has not been simulated.
However, given the use of an uncoated dielectric tip, dipole-dipole interactions between the tip and antenna are significantly reduced, supporting the comparison to the simulation conditions.

Qualitatively, there is agreement between the mode shape along the waveguide nanowires for the experimental and simulated near fields. 
Particularly in the asymmetric brightness of the bottom-right end of the waveguide compared to the top-left. 
One discrepancy between the experiment and simulation is the significantly brighter top-right end of the dimer at offsets 0.40\%-24.7\%. 
This is likely due to interference between light scattered from the tip dipole and the dimer. 
At the top-right end of the dimer, both resonators emit in phase and constructively interfere, while the opposite occurs at the bottom-left end of the dimer. 
Such interference is a consequence of measuring with p-polarised excitation in reflection using a polarisable tip.\autocite{Buchner2019_SNOMtipInterfer}
The dielectric tip with a lower polarisability was chosen over a metallic tip to reduce this effect but has not completely removed it.

\begin{figure*}[ht!]
    \centering
    \includegraphics[width=0.8\linewidth]{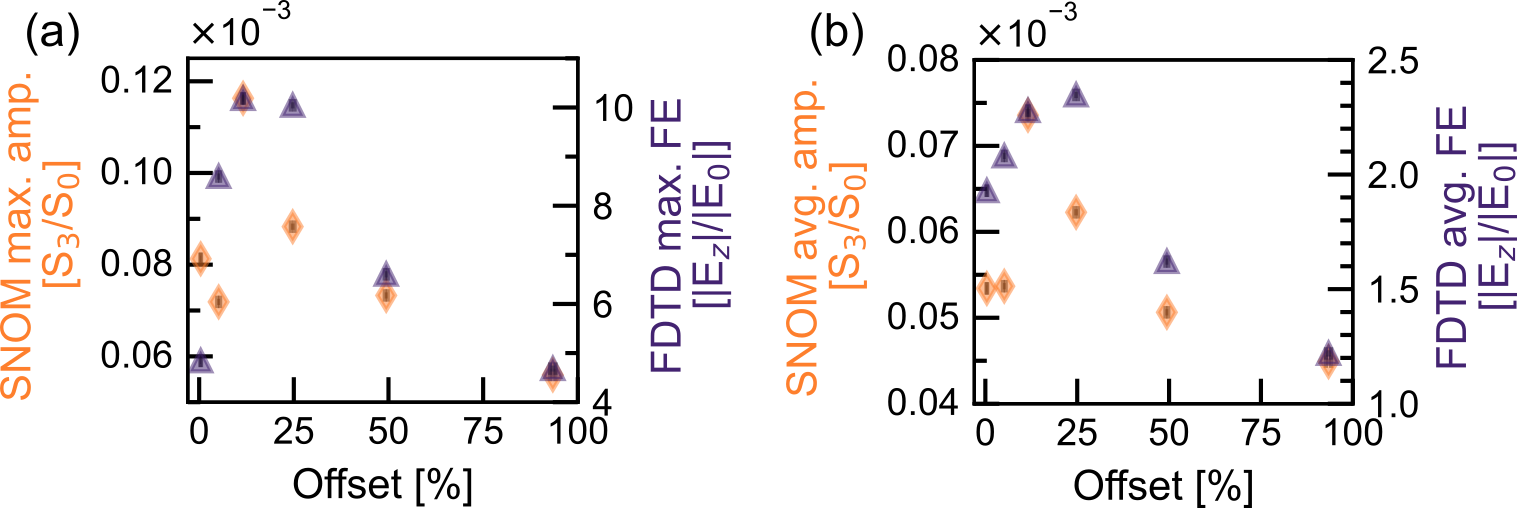}
    \caption{\textbf{Comparison between sSNOM amplitude (orange) and field enhancement (FE) of numerical simulations (FDTD, purple) on resonance}. 
    (a) Maximum field across the waveguide. 
    (b) Average field confined to the waveguide. Field analysed across the in-plane surface of the waveguide for all data.  }      
    \label{FIG_SNOMvsFDTD}
\end{figure*}

To quantify the amount of light reaching the waveguide mode, the maximum and average E-field confined to the in-plane surface of the waveguide in Fig.~\ref{FIG_CutWG_SNOM_amp} is assessed in Fig.~\ref{FIG_SNOMvsFDTD}. 
For the maximum field, the experiment and simulation peak at $11.6 \pm 0.2\%$ offset before falling to a minimum at $93.4 \pm 0.7\%$. 
For the average field, the simulations peak at $24.7\pm0.5\%$ while the experimental results peak at $11.6\pm 0.2\%$. 
The experimental results at $24.7\pm 0.5\%$ and $5.15\pm0.39\%$ deviate most from the trend suggested by the simulations. 
It is possible that the individual structures measured are defective due to fabrication errors and are not representative of their respective metasurfaces.

\begin{figure*}[ht!]
    \centering
    \includegraphics[width=0.925\linewidth]{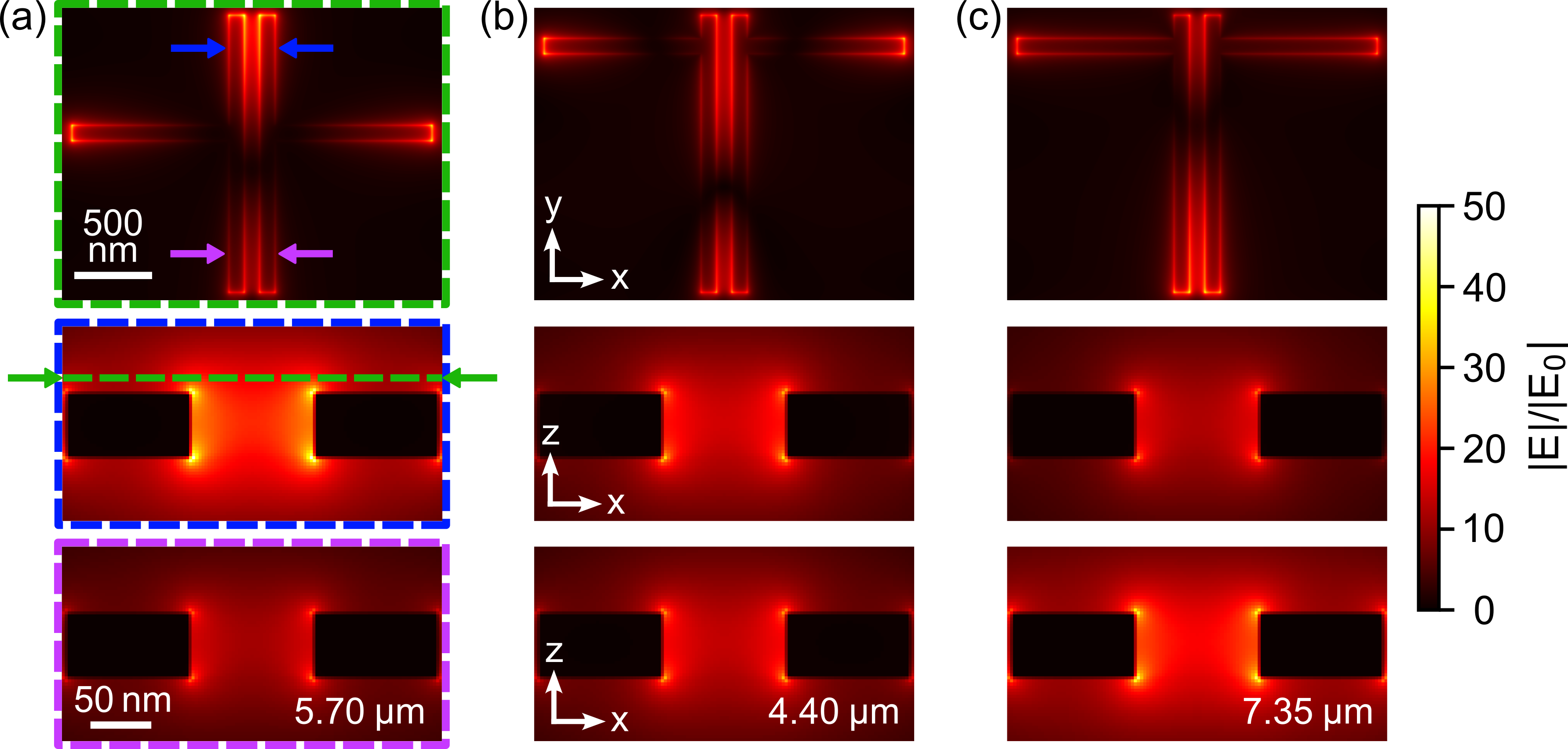}
    \caption{\textbf{Numerical simulations (FDTD) of the near-field modes for the nanostructures resonant at 5.7 \textmu m wavelength}. 
    (a) In- and out-of-plane profiles of the E-field mode for a 15\% offset nanostructure at a 5.7 \textmu m driving field. 
    In-plane data (green dashed box) was simulated 5 nm above the gold surface (green arrows). 
    Out-of-plane profiles (blue and purple dashed boxes) were recorded at cross-sections through the waveguide (blue and purple arrows). 
    (b) The higher energy split mode at 78\% offset and 4.40 \textmu m wavelength driving field. 
    (c) Lower energy split mode at 78\% offset and 7.35 \textmu m wavelength driving field. 
    The labelled axis and colour coded markings apply to all panels. 
    A 2.5 nm unit cell mesh has been used for all simulations.   }
    \label{FIG_CutWG_FDTD_modes}
\end{figure*}

As the sSNOM system operates in reflection, we cannot excite the nanostructures at normal incidence while collecting the near-field response. 
Instead, the simulation environment used to model the dimer-coupled waveguide system in Fig.~\ref{FIG_CutWG_SNOM_amp}(g-l) is rerun with excitation at normal incidence.
The result is the near-field amplitude modes in Fig.~\ref{FIG_CutWG_FDTD_modes}, which show the three-dimensional E-field components across the entire nanostructure in-plane and the waveguide mode out-of-plane.
At 15\% offset and a driving wavelength of 5.7 \textmu m, Fig.~\ref{FIG_CutWG_FDTD_modes}(a) represents the system when the average field is maximised in the waveguide but before the mode bifurcates.
The distribution of the E-field amplitude suggests a fundamental resonant mode, but with a slight asymmetry in the anti-node amplitude and node position. 
At a 78\% offset and the wavelengths of the hybrid resonances at (b) 4.40 \textmu m and (c) 7.35 \textmu m, the in-plane modes become more asymmetric regarding the node position, but in opposite directions along the waveguide.
As already suggested by Fig.~\ref{Fig_SI_CutWG_FDTD}(c), the amplitude in the waveguide is lower in the split modes compared to 5.7 \textmu m in the weak coupling regime. 
However, reasonably high field enhancements are still observed.
At both in-plane positions, the out-of-plane modes maintain a similar shape in the weak and strong coupling regimes.
Predictably, the greatest confinement is found at the gold-air interface, although a moderate enhancement is observed in the volume of the 100 nm air gap. 

In combination with the experimentally measured and simulated near field at angled excitation, these simulations verify that the waveguide mode of the fabricated dimer-coupled waveguides can be accessed at normal incidence.
It is assumed that the nanostructures resonant at 10 \textmu m wavelength have similar near-field modes, as the coupling mechanism remains identical.

\section{Coupled mode theory} \label{Apx_subsec_3lvl_scat_amp}

\subsection{Scattering amplitude of the two-resonator system}
Polarisation of a plasmonic structure is the collective displacement of electrons that can be modelled as a harmonic oscillator. The coupled plasmonic dimer antenna and resonant waveguide design is comprised of two polarisable structures coupled in the near field. Simplifying to one dimension without loss of generality, the system can be described by simple harmonic oscillator equations.
\begin{equation} \label{Apx_EQ_HarmOs1}
\ddot{x}_1(t) + \gamma_1 \dot{x}_1(t) + \omega_{1}^2 x_1(t) + \tilde{\kappa} \dot{x}_2(t) = g_1 E(t)
\end{equation}
\begin{equation} \label{Apx_EQ_HarmOs2}
\ddot{x}_2(t) + \gamma_2 \dot{x}_2(t) + \omega_{2}^2 x_2(t) - \tilde{\kappa} \dot{x}_1(t) = g_2 E(t)
\end{equation}
Here, $x_{1,2}$ are the displacements of the oscillators, $\gamma_{1,2}$ are the respective loss terms, $\omega_{1,2}$ are the resonant frequencies, and $\tilde{\kappa} = \kappa e^{i\phi_{\kappa}}$ is the complex coupling rate with amplitude $\kappa$ and phase $\phi_{\kappa}$. 
Considering the phase component of the coupling accounts for the retardation between the two resonators.\autocite{Taubert_2013_complexPhi_CMtheory} 
$E(t)$ is the external driving field and $g_{1,2}$ are the coupling coefficients of each resonator to this driving field. 
As the waveguide mode is optically dark to the far field, $g_{2}=0$ and $g_1=g$. 
The resonant frequencies of the system are designed to be degenerate and are rewritten as $\omega_0=\omega_1=\omega_2-\delta$, where $\delta$ accounts for any spectral detuning. 
Assuming a steady-state solution for the displacements and driving field -- \(x_{1,2}(t) =  x_{01,02}e^{-i \omega t}\) and \( E(t) = E_0 e^{-i \omega t}\), where \(\omega\) is the driving frequency and \(x_{01,02}\) are the resonator amplitudes -- we construct a coupling matrix equation.
\begin{equation}  
\label{EQ_matrix_CoupledHarm}
\begin{bmatrix}
\omega^2_{0}  - \omega^2- i \gamma_1 \omega  & -i\omega\widetilde{\kappa}  \\
i\omega\widetilde{\kappa} & (\omega_{0}+\delta)^2 - \omega^2 - i \gamma_2 \omega 
\end{bmatrix}
\cdot
\begin{bmatrix}
x_{01}   \\
x_{02} 
\end{bmatrix}
 =
\begin{bmatrix}
gE{_0}   \\
{0} 
\end{bmatrix}
\end{equation} 
The above matrix equation is identical to the coupled equations:
\begin{equation} \label{Apx_EQ_HarmOs_solu1}
(\omega_{0}^{2}-i\omega\gamma_1-\omega^2) x_{01} - i\omega \tilde{\kappa}x_{02} = g E_0  \mathrm{,}
\end{equation}
\begin{equation} \label{Apx_EQ_HarmOs_solu2}
((\omega_{0}+\delta)^{2}-i\omega\gamma_2-\omega^2) x_{02} + i\omega\tilde{\kappa} x_{01} = 0
\mathrm{.}
\end{equation}
Rearranging Eqn.~\ref{Apx_EQ_HarmOs_solu2} for the waveguide amplitude, $x_{02}$, substituting this into Eqn.~\ref{Apx_EQ_HarmOs_solu1}, and rearranging the result for $x_{01}$ allows the amplitude of the radiative dimer to be defined by all the oscillator parameters.
\begin{equation} 
x_{01} = \frac{gE_0}{\omega_0^2-i\omega\gamma_1-\omega^2 - \frac{(\omega \kappa e^{i\phi_{\kappa}})^2}{(\omega_0+\delta)^2-i\omega\gamma_2-\omega^2}}\label{Apx_EQ_HarmOs_x1Amp}
\end{equation}

The above equation represents a single compound nanostructure; however, the spectral measurements in the main text are taken across arrays of repeated structures that form metasurfaces. 
Eqns.~\ref{EQ_deAbajoMSscaling_r} and \ref{EQ_deAbajoMSscaling_t} account for this by considering dipole interactions in the near field and the collective far-field amplitude of multiple emitters.\autocite{deAbajo_MSscaling,Koenderink2020_MStransferMatrix}
\begin{equation}
r = \frac{2\pi ik}{\mathbb{A}}\frac{1}{\frac{1}{x_{01}} -\frac{2\pi ik}{\mathbb{A}}}   \label{EQ_deAbajoMSscaling_r}
\end{equation}

\begin{equation}
t = \frac{1}{1-\frac{2\pi ik}{\mathbb{A}} x_{01} } \label{EQ_deAbajoMSscaling_t}
\end{equation}

Here, $\mathbb{A}$ is the unit cell area of an array, measured from the SEM images of fabricated metasurfaces, and $t=1+r$. The intensities at the spectrometer detector are calculated as $R = |r|^2$ and $T=|t|^2$. Resultantly, spectral intensity can be represented by all coupling, loss, and resonance parameters of the plasmonic oscillators in Eqn.~\ref{EQ_matrix_CoupledHarm}. 

\subsection{Eigenvalues of the two-resonator system} \label{Apx_subsec_EV_EP_3lvl}
The system is driven close to resonance and the resonant peaks are narrow, validating the use of the rotating wave approximation and the simplification: \(\omega_{0}^2 - i \omega \gamma_{1,2} - \omega^2 \approx \omega_{0}^2 - i \omega_{0} \gamma_{1,2} - \omega_{0} \omega\). Applying this to Eqn.~\ref{EQ_matrix_CoupledHarm}, a simplified coupling matrix equation can be constructed to define the system under the assumption of \(\delta\approx0\).

\begin{equation}  
\label{Appen_Eq5_CoupledHarmSimp2}
\begin{bmatrix}
\omega_{0} - i \gamma_1 - \omega  & -i\kappa  \\
i\kappa & \omega_{0} - i \gamma_2 - \omega 
\end{bmatrix}
\cdot
\begin{bmatrix}
x_{01}   \\
x_{02} 
\end{bmatrix}
 =
\begin{bmatrix}
gE{_0}/\omega_0   \\
{0} 
\end{bmatrix}
\end{equation} 

\begin{figure*}[ht!]
    \centering
    \includegraphics[width=0.575\linewidth]{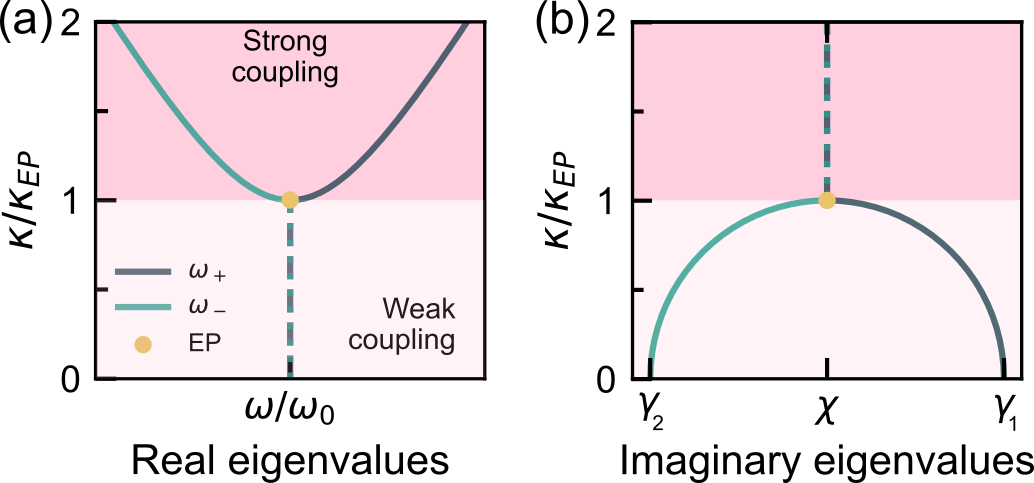}
    \caption{\textbf{The eigenvalues of the two-resonator system highlighting the strong and weak coupling regimes}. 
    (a) Real eigenvalues as a function of coupling strength, highlighting the transition from weak to strong coupling regimes. 
    (b) Imaginary eigenvalues as a function of coupling strength. }  \label{Fig_SI_2res_eigenvalues}

\end{figure*}

It should be noted that the coupling parameter is now purely real. Although the phase does account for spectral features,\autocite{Taubert_2013_complexPhi_CMtheory} it is typically found to be in the range \(0\leq\phi_\kappa\leq\pi/25\) and is dropped to simplify the following derivation.
The 2x2 matrix on the left side is a dressed Hamiltonian and solutions to Eqn.~\ref{Appen_Eq5_CoupledHarmSimp2} are found by taking its determinant.
\begin{equation} \label{Apx_Eq_Determinant}
(\omega_{0} - i \gamma_1 - \omega)( \omega_{0} - i \gamma_2 - \omega) - \kappa^2 = 0
\end{equation} 
Expanding and collecting like terms gives a quadratic equation  
\begin{equation} \label{Appen_Eq7_DetermLikeTerms}
\omega^2 - (2\omega_0 - i(\gamma_1 + \gamma_2))\omega + \omega_{0}^2 - i\omega_{0}(\gamma_1+\gamma_2) - \gamma_1 \gamma_2 - \kappa^2 = 0
\end{equation} 
\noindent with complex solutions
\begin{equation} \label{Apx_Eq_QuadSolu}
\omega_{\pm} = \omega_0 - i \chi \pm \sqrt{\kappa^2 - \beta^2} \mathrm{.}
\end{equation} 
Here, $\chi = (\gamma_1 + \gamma_2)/2$, $\beta = (\gamma_1 - \gamma_2)/2$ and the eigenvalues are degenerate at what is known as the exceptional point; $\kappa = \beta$.
The real and imaginary eigenvalues of Eqn.~\ref{Apx_Eq_QuadSolu} are visualised in Fig.~\ref{Fig_SI_2res_eigenvalues}, highlighting the weak and strong coupling regimes, as well as the exceptional point.

\subsection{Scattering amplitude of the three-resonator system} \label{Apx_subsec_molScat_amp}
The process used to derive the plasmonic scattering  amplitude above is applied to the three-resonator system, which includes a molecular resonance. 
An additional resonator is included in the coupling matrix that does not scatter to the far field or couple directly to the dimer, but does couple to the waveguide via the complex parameter $\tilde{\nu}=\nu e^{i\phi_{\nu}}$, where $\nu$ is the coupling amplitude and $\phi_\nu$ is the phase.
\begin{equation}
\begin{bmatrix}
  \omega^{2}_0 - i\omega\gamma_1 - \omega^{2} & -i\omega\tilde{\kappa} & 0 \\ 
  i\omega\tilde{\kappa} & (\omega_0 + \delta)^{2} - i\omega\gamma_2 - \omega^{2} & -i\omega\tilde{\nu} \\
  0 & i\omega\tilde{\nu} & \omega^{2}_m - i\omega\gamma_m - \omega^2
\end{bmatrix}
\cdot
\begin{bmatrix}
  x_{01}  \\
  x_{02} \\
  x_{0m} \\
\end{bmatrix}
=
\begin{bmatrix}
  gE_0  \\
  0 \\
  0 \\
\end{bmatrix} \label{CM_MolecularCouplingMatrix}
\end{equation} 

This matrix represents coupled simultaneous equations.
\begin{equation} 
x_{01}(\omega_0^2-i\omega\gamma-\omega^2)-x_{02}i\omega\tilde{\kappa} = gE_0
\label{Apx_EQ_moleCoup1}
\end{equation}
\begin{equation}
x_{01}i\omega\tilde{\kappa} + x_{02}((\omega_0+\delta)-i\omega\gamma_2 -\omega^2)-x_{0m}i\omega\tilde{\nu} = 0 
 \label{Apx_EQ_moleCoup2}
\end{equation}
\begin{equation} 
x_{0m} = -x_{02}\frac{i\omega\tilde{\nu}}{\omega_m^2-i\omega\gamma_m-\omega^2} \label{Apx_EQ_moleCoup3}
\end{equation}

Eqn.~\ref{Apx_EQ_moleCoup3} is substituted into Eqn.~\ref{Apx_EQ_moleCoup2} and rearranged for $x_{02}$.

\begin{equation} \label{Apx_EQ_moleCoup2_rearrange}
x_{02}=-x_{01}\frac{i\omega\tilde{\kappa}}{(\omega_0+\delta)^2 -i\omega\gamma_2-\omega^2 -\frac{\omega^2\tilde{\nu}^2}{\omega_m^2-i\omega\gamma_m-\omega^2}}
\end{equation}

Eqn.~\ref{Apx_EQ_moleCoup2_rearrange} is subsequently substituted into Eqn.~\ref{Apx_EQ_moleCoup1} and rearranged for the bright resonator amplitude $x_{01}$.

\begin{equation} \label{Apx_EQ_moleCoup1_rearrange}
x_{01}=\frac{gE_0}{\omega_0-i\omega\gamma_1-\omega^2 - \frac{\omega^2\tilde{\kappa}^2}{(\omega_0+\delta)^2-i\omega\gamma_2-\omega^2 - \frac{\omega^2\tilde{\nu}^2}{\omega_m^2-i\omega\gamma_m-\omega^2}}}
\end{equation}

Here, a single molecular resonance is assumed at \(\omega_m\). Additional resonances can be included in Eqn.~\ref{Apx_EQ_moleCoup3} by treating the molecular resonances as multiple Lorentzian oscillators that are uncoupled from one another.
\begin{equation} 
x_{0m} = -x_{02} \sum_{k} \frac{i\omega\tilde{\nu}}{\omega^2_{mk}-i\omega\gamma_{mk}-\omega^2} \label{Apx_EQ_mole_unCoup_sum}
\end{equation}

For the above equation, identical coupling strength to each spectral mode is assumed. Despite this limitation, the model accurately describes the experimental spectra in the main text.

\subsection{Eigenvalues of the three-resonator system} \label{Apx_subsec_EV_EP_4lvl}

Assuming spectral overlap between one of the hybridised plasmonic modes in the strong coupling regime and a single molecular resonance, \(\omega_\pm\approx\omega_m\), the three-resonator dressed Hamiltonian can be reduced to a 2x2 matrix by treating the plasmonic components as a single resonator with combined loss \(\chi\). 
\begin{equation}  
\label{Apx_4lvl_redu_Ham}
\begin{bmatrix}
\omega_{\pm} - i \chi - \omega  & -i\nu  \\
i\nu & \omega_{m} - i \gamma_m - \omega 
\end{bmatrix}
\end{equation} 

Again, only the amplitude of the coupling terms is considered here, due to \(\phi_{\nu}\approx0\). Following the same formalism as in section \ref{Apx_subsec_EV_EP_3lvl}, the determinant of the coupling matrix is calculated to reach complex solutions.
\begin{equation} \label{Apx_Eq_QuadSolu_mole}
\omega_{m\pm} = \omega_{\pm} - i \chi_m \pm \sqrt{\nu^2 -\beta^2_m}
\end{equation} 

Here, $\chi_m=(\chi+\gamma_m)/2$ and $\beta_m=(\chi-\gamma_m)/2$. The above equation is technically invariant to changes in $\kappa$ when $\kappa\geq\beta$) provided $\omega_0\approx\omega_m$. However, the spectral plasmonic hybrid modes shift significantly with $\kappa$, unlike the analogous dimer mode.

\section{CFW analysis}
To demonstrate how a complex driving frequency can compensate for the loss in a system, we introduce the scattering amplitude for a pair of one-dimensional uncoupled Lorentzian oscillators.
\begin{equation}
    x(\omega^\prime) = E_0 \left( \frac{1}{\omega_1^2-\omega^{\prime2}-i\gamma\omega^{\prime}}+\frac{1}{\omega_2^2-\omega^{\prime2}-i\gamma\omega^{\prime}}\right)
\end{equation}

\begin{figure*}[ht!]
    \centering
    \includegraphics[width=0.6\linewidth]{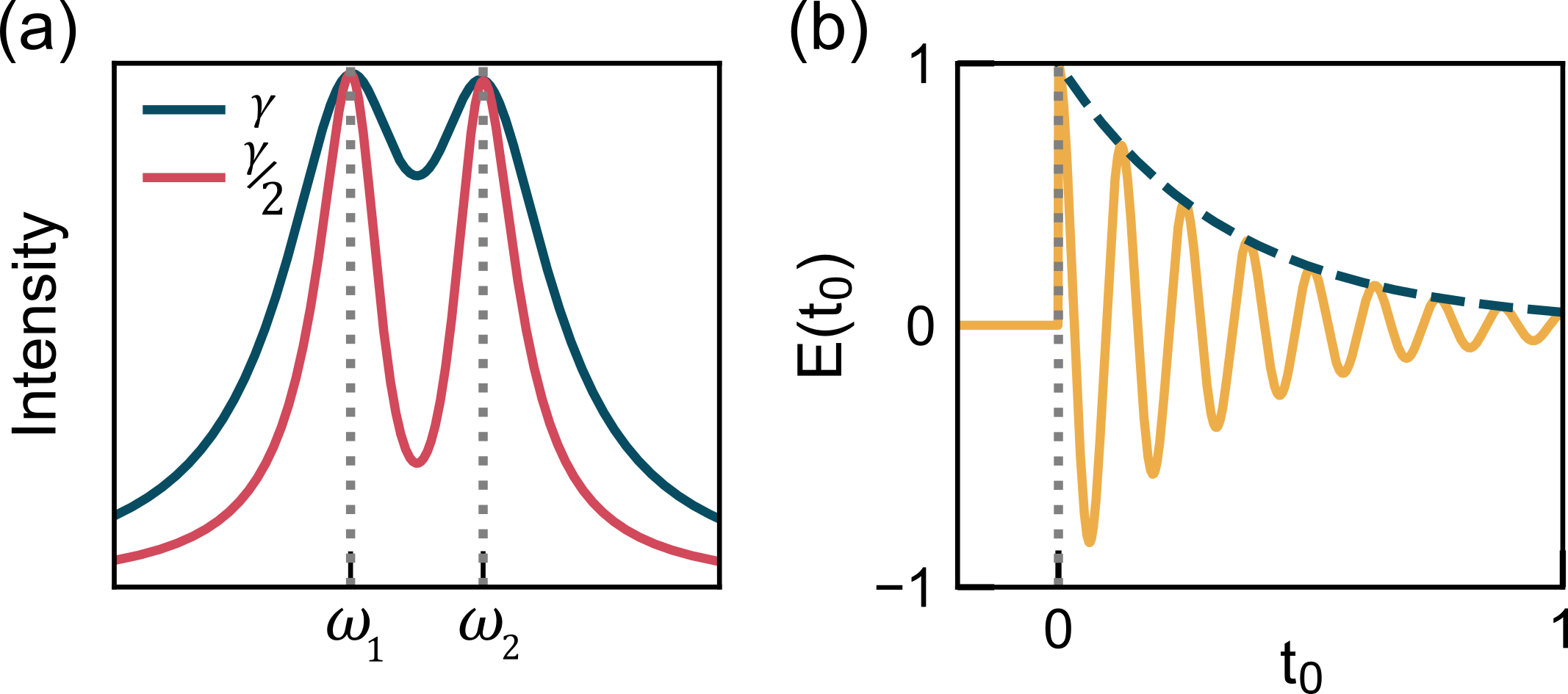}
    \caption{\textbf{The principles of CFW analysis}. (a) Scattering intensity of a system of two uncoupled Lorentzian oscillators before and after loss is reduced. (b) Amplitude of an attenuated and truncated CFW. The dark blue dashed line indicates the attenuation due to the imaginary component, $\tau$.  }
   
    \label{fig1_CFWintro}
\end{figure*}

$E_0$ is the perfectly coupled amplitude of the driving field at frequency $\omega$. The detuning between the resonant frequencies $\omega_1$ and $\omega_2$ is on the order of their shared loss rate, $\gamma$, such that the corresponding peaks in the spectral intensity in Fig.~\ref{fig1_CFWintro}(a) are poorly resolved.
Driving the system at a complex frequency $\widetilde{\omega} = \omega^\prime-i\tau/2$ leads to the amplitude 
\begin{equation}
    x(\widetilde{\omega}) = E_0 \left( \frac{1}{\omega_1^2-\omega^{\prime2}-(\frac{\tau\gamma}{2}-\frac{\tau^2}{4})-i\omega^\prime(\gamma- \tau)} + \frac{1}{\omega_2^2-\omega^{\prime2}-(\frac{\tau\gamma}{2}-\frac{\tau^2}{4})-i\omega^\prime(\gamma- \tau)}\right) \mathrm{.}
\end{equation}

Upon comparison of the two scattering amplitude equations, it can be seen that the imaginary component of the driving frequency reduces the loss rate of the oscillators, $\gamma-\tau$, potentially compensating for the loss entirely. 
More conservatively, $\tau=\gamma/2$ significantly improves the spectral resolution of the resonant peaks in Fig.~\ref{fig1_CFWintro}(a).
The complex illumination also introduces an additional real term $\frac{\tau\gamma}{2}-\frac{\tau^2}{4}$ that produces a shift in the observed resonant frequency. 
For typical nanophotonics systems with comparatively lower loss to resonant frequencies, $\gamma \ll \omega_{1,2}$, this shift is small. 
For reference, $\gamma/\omega_{1,2}$ and $(\omega_2-\omega_1)/\omega_{1,2}$ are both $\sim10^{-2}$ in Fig.~\ref{fig1_CFWintro}(a). 
A propagating wave in free-space with a complex frequency $\widetilde{\omega}$ experiences attenuation over time, where $\tau$ is the attenuation factor. 
Illuminating with such a CFW poses practical difficulties; particularly as the strength of the attenuation coefficient defines the degree of loss compensation.

Fortunately, synthetic CFWs allow retro-active excitation of a system to produce a similar effect.
To achieve this, a synthetic CFW is defined as $E(t_0) = E_0e^{-i\widetilde{\omega}t_0}\theta(t_0)$, where $\theta(t_0)$ is a truncation in time of $\theta(t_0)=0$ for $t_0<0$ and $\theta(t_0)=1$ for $t_0\geq0$. 
This truncation is necessary to prevent unphysical divergence at negative time and results in the waveform in Fig.~\ref{fig1_CFWintro}(b). 
By taking the Fourier transform of the CFW, it can be rewritten as
\begin{equation}
E(t_0) = \theta(t_0) \frac{E_0}{2\pi} \int^\infty_{-\infty} \frac{1}{i(\widetilde{\omega}-\omega)} e^{-i\omega t_0} d\omega \mathrm{,}
\label{EQ_CFW_FT}
\end{equation} 
where the Fourier coefficient is used to apply a frequency domain convolution to a general scattering amplitude $\widetilde{P}(\omega)$.
\begin{equation}
P(\widetilde{\omega}) \approx  \int^\infty_{-\infty} \widetilde{P}(\omega) \frac{1}{i(\widetilde{\omega}-\omega)} e^{i(\widetilde{\omega}-\omega) t_0} \frac{d\omega}{2 \pi}
\label{EQ_CFW_freqDomainconvol}
\end{equation}
The result is a scattering amplitude as a function of complex-frequencies.
Experimental data is typically discrete; therefore, the convolution is discretised across real frequency points.
\begin{equation}
    P(\widetilde{\omega}) \approx \frac{1}{2\pi} \sum_{n} \widetilde{P}(\omega_n)\frac{1}{i(\widetilde{\omega}-\omega_n)} e^{i(\widetilde{\omega}-\omega_n)t_0} \Delta\omega
    \label{EQ_CFW_response_FT_discrete}
\end{equation}
This is a coherent summation across real frequency points and requires amplitude and phase information of the initial response, $\widetilde{P}(\omega_n)=P(\omega_n)e^{i\phi(\omega_n)}$.
Phase information is not typically collected alongside spectral intensity, $I(\omega_n)$, but can be calculated using the discretised Kramers-Kronig relation.
\begin{equation}
    \phi(\omega_n) = \frac{1}{2\pi} \sum_{n} \frac{\ln{|I(\omega_n)}|}{\omega-\omega_n} \Delta\omega
    \label{EQ_CFW_KK_discrete}
\end{equation}
In order for Eqn.~\ref{EQ_CFW_KK_discrete} to converge,
$\ln{|I(\omega)|} \rightarrow 0$ when $\omega \rightarrow \infty$, requiring $I(\omega) \rightarrow 1$. 
This is generally not the reality for experimental measurements, and the spectral region of interest typically needs to be extrapolated to $1$. 
A spectral response that tends to zero must be transformed to meet this requirement.

$\widetilde{P}(\omega)$ can then be convoluted by Eqn.~\ref{EQ_CFW_response_FT_discrete} to determine the scattering response to the synthetic CFW, $P(\widetilde{\omega})$. 
From $P(\widetilde{\omega})$, the corresponding transmission and reflection amplitudes are calculated using $r(\widetilde{\omega})=-\frac{P(\widetilde{\omega})}{P(\widetilde{\omega})+i}$ and $t(\widetilde{\omega})=\frac{i}{P(\widetilde{\omega})+i}$, for comparison with the unprocessed measurement.

The truncation in Eqn.~\ref{EQ_CFW_FT} prevents the system from immediately reaching a quasi-steady state at $t=0$ and instead approaches it over time. 
Consequently, unphysical sidebands appear in the processed spectra, which are combatted by averaging over a time window, \(0\leq t_0 \leq t_{max}\).\autocite{Zeng2024_CFW_MoleSens} 
Unfortunately, this averaging cannot be performed indefinitely due to $E(t_0)\approx0$ at shorter $t_0$ as $\tau$ is increased.
Synthetically illuminating with near zero amplitude causes the process to diverge, meaning an appropriate $t_{max}$ must be chosen for each value of $\tau$.

\section{Plasmonic parameter comparison}
\begin{figure*}[ht!]
    \centering
    \includegraphics[width=0.8\linewidth]{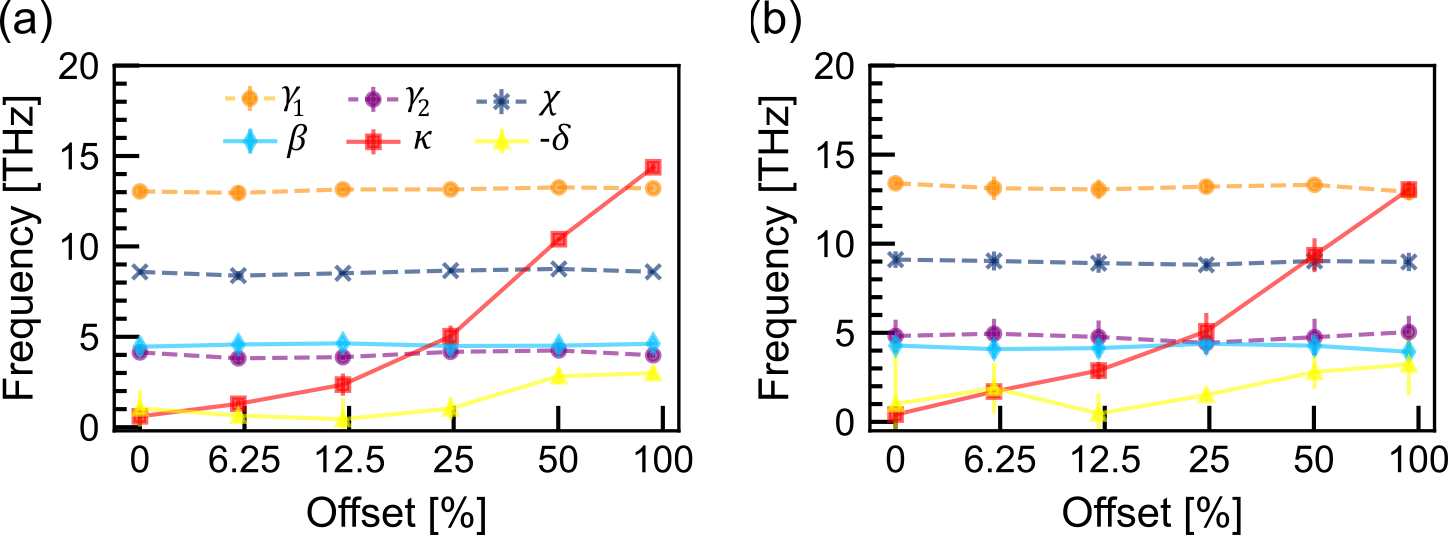}
    \caption{\textbf{The coupling and resonance parameters for the plasmonic resonators extracted using the two- and three-resonator models}. (a) Plasmonic parameters extracted from uncoated spectral data using the coupled-mode theory for two resonators. (b) Plasmonic parameters extracted from BPT coated spectral data using the coupled-mode theory for three resonators.   } 
    \label{Fig_SI_plasParamComp}
\end{figure*}

The coupled-mode theory for three resonators has a large parameter space, which makes the fit to experimental spectra liable to converge to unphysical values without strict bounds.
To prevent this, it is assumed that the plasmonic coupling and resonance parameters remain unchanged by the BPT monolayer, and the extracted uncoated parameters are used to define bounds for the three-resonator model.
The plasmonic parameters extracted from the uncoated and BPT coated spectra, respectively, are shown in Fig.~\ref{Fig_SI_plasParamComp}.
The extracted plasmonic parameters from each model are largely in agreement; although the waveguide loss is marginally higher and coupling at the largest two offsets has fallen slightly after coating. 
Overall, combined with the fittings shown in Fig.~4(c) and (d) of the main text, we are confident that realistic physical parameters are reached for all resonators and that the spectral response is modelled accurately.

\section{Identifying molecular modes}
\begin{figure*}[ht!]
    \centering
    \includegraphics[width=0.75\linewidth]{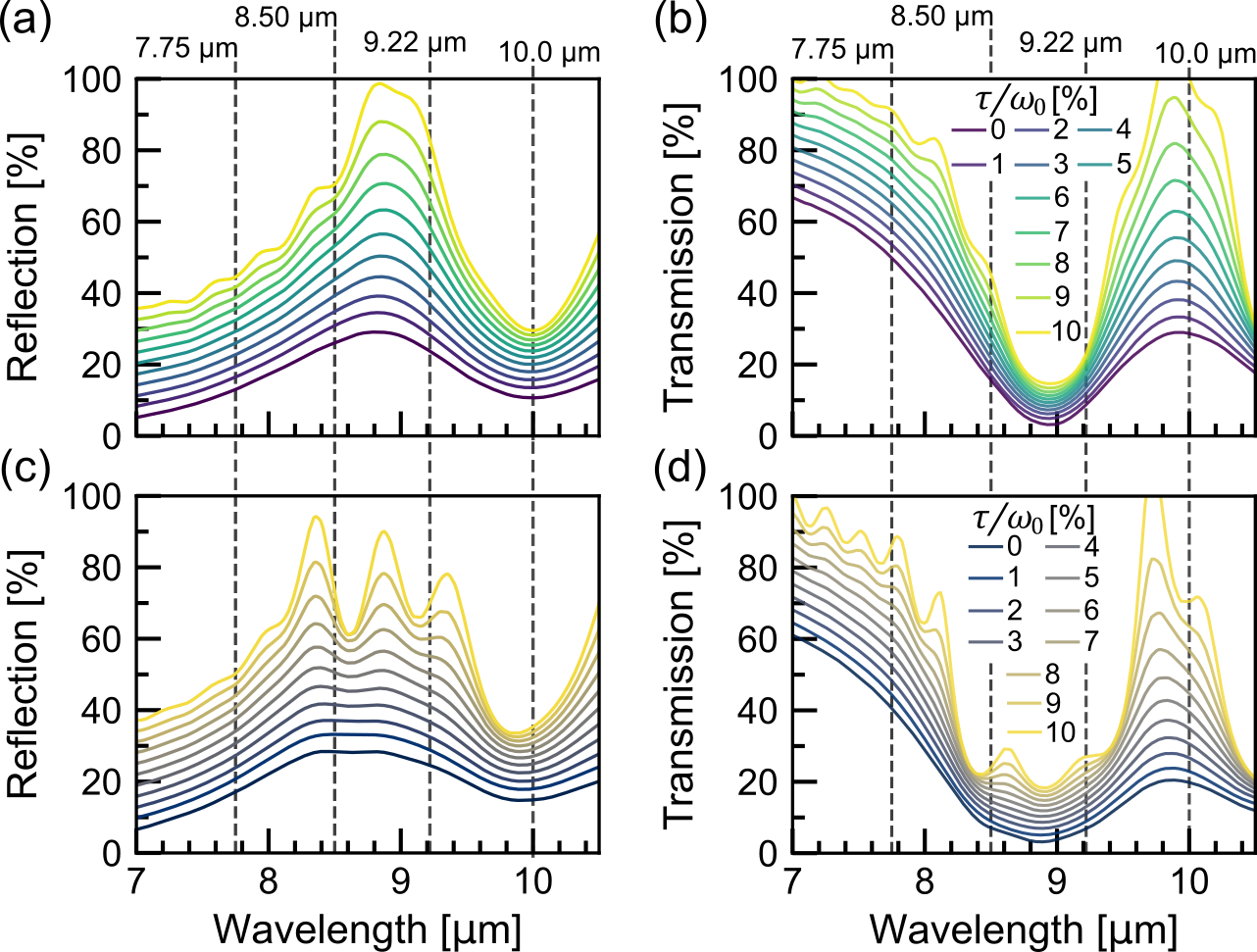}
    \caption{\textbf{The spectral response of the 25.1\% offset metasurface at various virtual gain values before and after coating with BPT}. (a,b) Reflection and transmission of the uncoated metasurface (c,d) Reflection and transmission of the BPT coated metasurface. The spectra spans the range where IR BPT resonances are expected.}
   
    \label{Fig_SI_CFW_UnC_Compare}
\end{figure*}

Identifying spectral modes that are initially completely obscured by loss is challenging due to the unphysical sidebands that appear at high virtual gain values.
In order to differentiate genuine modes from sidebands, Fig.~\ref{Fig_SI_CFW_UnC_Compare} compares reflection and transmission intensity at a range of virtual gain, before and after coating with BPT.
The sidebands begin to appear from $\tau/\omega_0\ge7\%$;
however, their spectral positions are similar whether BPT is present or not. 
This allows the resonant modes that appear after BPT has been coated to be identified at $8.5$ and $9.22$ \textmu m. 
There is no clear change after coating for any virtual gain value at $7.75$ \textmu m, and sidebands are prominent in this region.

The $10.0$ \textmu m resonance is difficult to identify with certainty.
There is a significant difference in transmission between the uncoated and BPT coated resonance at $10.0$ \textmu m from $\tau/\omega_0\ge7\%$, but the potential mode in Fig.~\ref{Fig_SI_CFW_UnC_Compare}(d) has a prominence on the same scale as the sidebands.
This is because the sidebands are more prominent in transmission after coating than in any of the other spectra.
Conversely, in reflection, the shift in the dip observed at $10.0$ \textmu m after coating suggests an IR mode is present.
However, because the $10.0$ \textmu m resonance of BPT cannot be definitively identified, the analysis performed in the main text models two resonances at $8.5$ and $9.22$ \textmu m.
Furthermore, these results concur that the upper limit of virtual gain that can be applied to the system is $\tau/\omega_0=6\%$.

\printbibliography